\documentclass[aps,prb,reprint,groupedaddress,floatfix]{revtex4-2}

\usepackage[utf8]{inputenc} 

\usepackage{geometry} 
\usepackage{graphicx} 
\usepackage[graphicx]{realboxes}

\usepackage{booktabs} 
\usepackage{array}    
\usepackage{paralist} 
\usepackage{verbatim} 
\usepackage{subfig}   
\usepackage{tabularx}
\usepackage{footnote}
\usepackage{tablefootnote}
\usepackage{multirow}
\usepackage{color,soul}

\usepackage{fancyhdr} 
\usepackage{times}

\usepackage{ulem} 
\usepackage{graphicx}
\def\equationautorefname#1#2\null{Eq.#1(#2\null)}

\usepackage{amsmath}
\usepackage{amssymb}
\usepackage{amsthm}
\usepackage{hyperref}
\usepackage{cancel}
\usepackage{IEEEtrantools}
\newcommand{\ii}{\mathrm{i}}
\newcommand{\ZZ}{\mathbb{Z}}
\newcommand{\Qt}{\textsf{Q}_\textsf{8}}
\newcommand{\Dn}{\textsf{D}_\textsf{8}}

\newcommand{\xb}{\overline{x}}
\newcommand{\yb}{\overline{y}}
\newcommand{\zb}{\overline{z}}

\newcommand{\kk}{\mathbf{k}}
\newcommand{\Aut}{\textrm{Aut}}
\newcommand{\oh}{\frac{1}{2}}
\newcommand{\oq}{\frac{1}{4}}
\newcommand{\tm}{\frac{3}{2}}
\newcommand{\tq}{\frac{3}{4}}
\newcommand{\x}{\times}

\newcommand{\zz}{\mathbf{0}}
\newcommand{\ie}{\textit{i.e. }}
\begin{document}

\title{Local symmetry groups for arbitrary wavevectors}
\author{Emanuele Maggio$^{\textrm{1}}$}
\thanks{emanuele.maggio@gmail.com}~
\author{~Andriy Smolyanyuk$^{\textrm{1}}$}
\author{~Jan M. Tomczak$^{\textrm{1,2}}$}
\affiliation{$^{\textrm{1}}$Institute of Solid State Physics, Technische Universit{\"a}t Wien,  Wiedner Hauptstra{\ss}e 8-10, A-1040 Vienna, Austria}

\affiliation{$^{\textrm{2}}$Department of Physics, King’s College London, WC2R 2LS, Strand, London, United Kingdom}

\begin{abstract}
We present an algorithm for the determination of the local symmetry group for arbitrary $\kk$-points in 3D Brillouin zones.
First, we test our implementation against tabulated results available for standard high-symmetry points (given by universal fractional coordinates). Then, to showcase the general applicability of our methodology, we produce
the irreducible representations for the ``non-universal high-symmetry"  points, first reported by Setyawan and Curtarolo [Comput.\ Mater.\ Sci.\ 49, 299 (2010)].
The present method can be regarded as a first step for the  determination of elementary band decompositions and symmetry-enforced constraints in crystalline topological materials. 
\end{abstract}

\maketitle

\section{Introduction}
\label{sec:Intro}

	Topological materials have entered the centre stage of both theoretical \cite{Hasan2010, Burkov2016, Bansil2016, Chiu2016, Cano2020}
	and experimental \cite{Chen2009, Xu2012, Liu2014, Kung2019, Kumar2019}
	investigations in condensed matter physics, thanks to possible applications they offer \cite{Kitaev2003, Nayak2008, Bandres2018, Politano2018} and to the vast array of conceptual challenges they pose.
	Limiting the discussion to theoretical aspects,  we can summarise the most fundamental question as follows: given a specific material (\ie a crystalline solid with a definite chemical composition and stoichiometry) and its (weakly interacting) band structure, is it possible to continuously connect it to an atomic limit? 
	In other words: when can one represent the electronic structure of the material in terms of its chemical constituents and when do single particle dispersions emerge that define globally a (topologically) non-trivial state of matter?
	
	Generalising the original approach by Dyson \cite{Dyson1962}, efforts for classifying systems of free electrons with and without time-reversal symmetry have come to fruition for systems without discrete spatial symmetries \cite{Kitaev2009}.
	Towards a more general classification, discrete spatial symmetries pose considerable difficulties to incorporate in full generality \cite{Shiozaki2022}, 
	yet, they appear to protect the topological nature of systems featuring an inversion centre \cite{Fu2007},  mirror reflections \cite{Teo2008, Hsieh2012} or non-symmorphic symmetry elements \cite{Shiozaki2015,Fang2015a,Gomi2019,Kim2019b,Yoshida2019a}.
		
	Further, to study topological semimetals, the concept of band crossing has been often used, which implies identifying a certain symmetry (that must be preserved along the energy band) and any obstruction of the band connectivity at high symmetry $\kk$-points in a way that preserves said symmetry \cite{Young2012, Steinberg2014}. 
	On the other hand, for the class of Weyl semimetals, the application of symmetry criteria is problematic, since, owing to the degeneracies of the electronic wavefunction with spin-orbit coupling, such band crossings can occur at any point in reciprocal space \cite{Song2018a}.
	Recently, in chiral crystals, the interplay  at high symmetry points between structural symmetries and time inversion has been investigated \cite{Chang2018}, thus allowing the identification of symmetry-enforced band crossings also for these systems \cite{Hirschmann2021}.
	Additionally, a connection has been highlighted between topological invariants, which correspond to certain values of symmetry indicators, and Berry phases evaluated over closed loops in the Brillouin zone \cite{Song2018a}.
	
	Systematic approaches to the identification of symmetry indicators for topological materials have been developed recently	\cite{Bradlyn2017,Po2017,Song2018,Khalaf2018,
	Yu2021,Tang2018a,Zhang2018}:
	a common theme of these methods is how the materials' symmetries are taken into account through their action on the single particle wavefunctions and the corresponding energy dispersion bands.
	Specifically, the overarching rationale for identifying a topological material is through an obstruction of the decomposition of energy band dispersions in reciprocal space  into representations of the (topologically trivial) atomic insulator, that is, to determine its elementary band decomposition.
	This approach banks on a seminal idea, due to Zak \cite{Zak1981,Zak1982b}, where space group representations are induced starting from either those of high symmetry points in reciprocal space, or Wyckoff positions in direct space.
	For a topologically trivial material, it is surely possible to decompose one such representation in terms of the elementary bands, and an obstruction to do so, identifies a topological band structure.

	The key ingredient for either the identification of symmetry indicators or the decomposition into elementary bands is the local symmetry group of the $\kk$-points considered, \mbox{a.\ k.\ a.\ }the little group of the wavevector.
	These groups have long been tabulated \cite{Bradley1972a} for a conventional set of $\kk$-points, and recently \cite{Liu2020a} a digitalised version of the old tables has been produced.
	The same tables are also accessible on the \mbox{Bilbao Crystallographic Server \cite{Aroyo2006}}.
	
\begin{figure}
	\includegraphics[width=70mm]{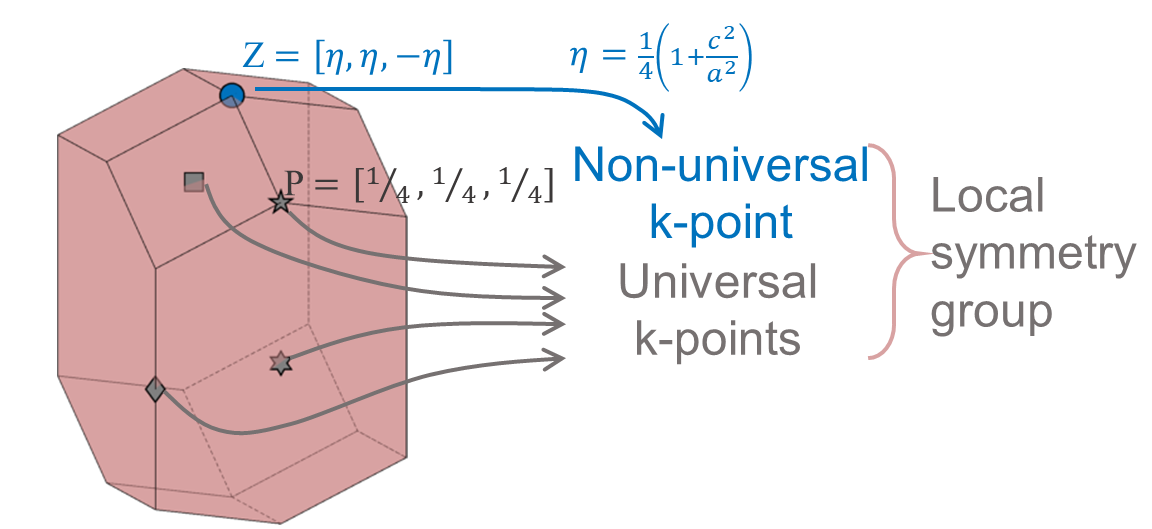}
	\caption{Schematic representation of the classification for $\kk$-points on the Brillouin zone boundary. Here we provide a consistent implementation for the two cases depicted.}
	\label{fig:GraphAbs}
\end{figure}

	In a detailed study Setyawan and Curtarolo \cite{Setyawan2010} report additional high symmetry $\kk$-points \footnote{we adopt this nomenclature to be consistent with the previous literature, however it is a bit of a misnomer as the $\kk$-points themselves do not have higher symmetry than all neighbouring points, rather, they are mid- or end- points of high symmetry lines} necessary to identify continuous dispersion paths in reciprocal space.
	More recent efforts \cite{Hinuma2017,Munro2020} have further relaxed the definition for the high symmetry $\kk$-points and provided a common frame for different crystallographic conventions.
	For all the newly tabulated $\kk$-points, the coordinates (reported in Figs. \ref{fig:mP}-\ref{fig:cF} in the primitive Wigner-Seitz cell) come to depend on the lattice parameters \cite{Setyawan2010,Hinuma2017}.
	We refer to these $\kk$-points as ``non-universal" to contrast them with the conventional set of high symmetry points for which the coordinates can be represented by universal fractions independent of the lattice parameters.
	It must be pointed out that these ``non-universal" $\kk$-points have the same stabiliser group as the high symmetry line they occupy and they can be chosen as line representatives if also the line lies on the Brillouin zone boundary; on the other hand, if such ``non-universal" $\kk$-points are the only intersection with the Brillouin zone boundary, the corresponding local symmetry group will have to include additional translations that act trivially on $\kk$-points inside the Brillouin zone, hence resulting in a different local symmetry group than the rest of the high-symmetry line.
	In \autoref{fig:GraphAbs}, we schematically exemplify the distinction between ``non-universal" and universal $\kk$-points:
	while for the shown body-centred tetragonal Brillouin zone for lattice constants $a>c$, \textit{e.g.}, the P-point $[\frac{1}{4},\frac{1}{4},\frac{1}{4} ]$ is a universal rational momentum-coordinate, the ``non-universal" Z-point [$\eta$, $\eta$, $-\eta$] explicitly depends on the lattice parameters as $\eta=(1+c^2/a^2)/4$.
	
	In all the above approaches the high symmetry points are identified through the action of space group operations that leave it invariant (up to the addition of a reciprocal lattice vector), \ie via the $\kk$-point stabiliser  $G_\kk$.
	Yet, as we have already highlighted above, the key ingredient through which $\kk$-points enter the classification of electronic band structures is the irreducible representation of the little group of the wavevector, $G^*_\kk$.
	However, an evaluation of the latter poses a challenge for the usual method (see \autoref{ssec:ThSum} for a summary) at ``non-universal" $\kk$-points, since also the corresponding little group of the wavevector turns out to depend on structural details.
	On the Bilbao Crystallographic server this issue is avoided by reporting only the stabiliser group of the symmetry line, which implies that the irreducible representations will not contain the translational elements that act on the Brillouin zone boundary. The characters corresponding to these translations can be obtained easily in the case of a split extension, but they are less straightforward to evaluate for non-symmorphic space groups. 
	While neglecting them might suffice for most practical applications,
	in this work we outline and implement a theory that treats universal and ``non-universal'' $\kk$-points on an equal footing by proposing an algorithm that can identify the little group for an arbitrary $\kk$-point.
	Specifically, we apply the method to the lattice parameter-dependent high-symmetry points and present the character tables defining their irreducible representations; the explicit matrix representations are also reported in the supplementary information.

The paper is organized as follows:
	 we summarise in \autoref{ssec:ThSum} some background about the conventional method to construct $G^*_\kk$, and in \autoref{ssec:Aut} we instead devise a more general approach for identifying group extensions, without any dependence from the lattice parameters.
	 Our algorithm allows to consider all high symmetry points for the 230 space groups in 3 dimensions. 
	Details of our computational implementations are reported in \autoref{sec:comp} and we finally list the groups $G^*_\kk$ and their character tables in \autoref{sec:results}.

\section{Theory}
\label{sec:Th}

\subsection{Background}
	In this section we provide for completeness some background for the ensuing presentation in \autoref{ssec:Aut}, and we refer to Refs.~\cite{HCGT,Hiller1986} for a more detailed discussion.
	To ease the notation, we drop the subscript $\kk$ from the groups introduced previously, hence we deal in general with an Abelian group $M$ that is a normal subgroup of $G^*$, whereas $G \cong G^*/M$ need not be a subgroup of $G^*$. 
	We assume that there is a homomorphism $\pi: G^* \to G$ that makes the following sequence of groups
	\begin{align}
	\label{eq:ExtSeq}
	e \rightarrow M \rightarrow G^* \xrightarrow{\pi} G \rightarrow e
	\end{align}
	exact.
	That means that $\ker \{\pi \} = M$, hence we can find a transversal (\ie a set of coset representatives) for $G$ in $G^*$, $t: G \to G^*$, such that $t(e) = e$, with $e$ the identity element (of the appropriate group).  
	The elements of $G^*$ can then be written as $(t(x),m)$ for $x \in G$ and $m \in M$.
	For any two elements $x,y \in G$ one has $\pi(t(xy)) = xy = \pi(t(x)) \pi(t(y)) = \pi(t(x)t(y))$, hence there is a unique element $\mu(x,y) \in M$ such that 
	\begin{align}
	t(xy) \mu(x,y) = t(x)t(y).
	\label{eq:prd}
	\end{align}
	The elements $\mu$ of $M$ are \mbox{2-cocycles}, but also known as Schur's multipliers and represent the ``obstructions" in $G^*$ that do not allow $G$ to be a subgroup of $G^*$.

	Since the little group of $\kk$ (here: $G^*$) comprises of both point group operations and translations, its identification is equivalent to the construction of the correct group extension of $M$ by $G$, where $M$ can now be identified with the subgroup of lattice translations. 
	We find it more convenient to stick to the multiplicative notation for elements of $M$, instead of using the common additive notation.
	A defining feature of the group $G$ is to leave the lattice $M$ invariant,
	and if we indicate the action of an element $x \in G$ on $m \in M$ by $m^x = m^{t(x)}$, we have a homomorphism from $G$ to the group of transformations of $M$ into itself $\alpha: G \to \Aut(M), x \mapsto x^\alpha$, which will be required in \autoref{ssec:Aut}.
	
\subsection{Summary of the standard approach}
\label{ssec:ThSum}

	The method of choice in solid state physics for the construction of the little group of the wavevector is due to Herring \cite{Herring1942,El-Batanouny2008} and it has been applied to all 230 space groups in Ref.~\cite{Bradley1972a} for the restricted list of $\kk$-points with rational coordinates.
	
	The method produces an extension of the stabiliser of the $\kk$-point by introducing the map $s: M \to \mathbb{C}$, $s := \textrm{exp}[ 2\ii \pi \kk \cdot \mu]$ that effectively identifies non-trivial translations;
	 the value of $s$ is then adjoined to each element of $G$ to label the entries of the new group $G^*$, whose multiplication table is generated following the point group multiplication rule for the first index and the product rule in \autoref{eq:prd} with the definition above for the second index \cite{El-Batanouny2008}.
	The action of these translations on the point group operations is trivial for symmorphic space groups, meaning that the resulting little group of the wavevector is the direct product of the stabiliser with the cyclic group generated by the relevant translations. 
	For non-symmorphic space groups this may not be the case and $G^*$ is in general a group with a more complex structure, with non-trivial relations involving point group operations and translations.

	The group's order can also be determined from knowledge of the $\kk$-point alone \cite{Car1975}: $|G^*| = p |G|$ with $p = \textrm{lcm}(p_1,p_2,p_3)$ for a $\kk$-point whose coordinates are the rational numbers $\kk = [\frac{q_1}{p_1},\frac{q_2}{p_2},\frac{q_3}{p_3}]$.
	For the ``non-universal" $\kk$-points such an approach is clearly not viable, since the order of the resulting group would depend on the crystallographic lattice constants, rather than be uniquely determined by the symmetries at play for that particular wavevector.
	In the following section we suggest an alternative construction for the identification of the little group of the wavevector, thus filling a conspicuous gap in the recent literature.

\subsection{Central Extension's Automorphism group }
\label{ssec:Aut}
	
	To overcome the limitations stated above, we propose to leverage an idea due to Wells \cite{Wells1971,Passi2010} which allows to characterise the group extension $G^*$ by studying the transformations that map $G^*$ onto itself, that is by studying the group $\Aut( G^*)$.
	
	In particular, for transformations $\theta \in \Aut ( M)$ and $\phi \in \Aut( G)$, there might not be a transformation $\gamma \in \Aut ( G^*)$ that induces the pair $(\theta,\phi)$.
	A pair $(\theta,\phi) \in \Aut(M) \x \Aut(G)$ is compatible if it fulfils the condition $\theta x^\alpha \theta^{-1} = \left( x^\phi \right)^\alpha$ for all $x \in G$, in other words, $(\theta,\phi)$ are compatible if they preserve the conjugation action of $G$ on $M$.
	Compatible pairs form a group \cite{HCGT,Passi2010} $\mathsf{Comp}(G^*,M)$ contained in  $\Aut(M) \x \Aut(G)$. 
	So, if there is an automorphism $\gamma$ of $G^*$ that keeps $M$ fixed (as a set, so elements of $M$ could be permuted among each other) then one could define the map $\tau: \Aut_M(G^*) \to \Aut(M) \x \Aut(G)$ as $\tau(\gamma) = (\theta,\phi)$, which is crucial for the definition of the following exact sequence constructed by Wells \cite{Wells1971}:
\begin{widetext}	
	\begin{align}
	e \rightarrow Z^1_\alpha\left( G,M \right) \xrightarrow{\psi} \Aut_M \left( G^* \right) \xrightarrow{\tau} \mathsf{Comp}(G^*,M) \rightarrow H^2_\alpha(G,M)\rightarrow e.
	\label{eq:ExSeq2}
	\end{align}
\end{widetext}

	The sequence above connects the groups that we have just constructed with the second cohomology group $H^2$ and with the group of 1-cocycles $Z^1$.
	In particular, since the sequence is exact, one has that $ \mathrm{im}\{\psi\} = \ker\{\tau\}$, thus it sufficient to study the image and the kernel of $\tau$ to fully characterise the transformations of $G^*$ that we are interested in. 
	Since the sequence above is exact, these transformations are all mapped into the identity element of $H_\alpha^2$.
	
	More specifically, the following theorem allows to explicitly construct the \mbox{2-cocycles} that specify our little group of the wavevector \cite{Wells1971}:
	if $\gamma \in \Aut_M(G^*)$ then there is a triplet $(\theta,\phi,\chi) \in \Aut(M) \x \Aut(G) \x V/M$ such that
\begin{widetext}	
	\begin{align}
	&\gamma \left( \left(t(x), m \right) \right) = 
	\left( \phi(x), \chi(x)\theta(m)\right) 
	\label{eq:gamma} \\
	&\theta\left( m^x \right) = \theta(m)^{\phi(x)}
	\label{eq:compAction}\\
	&\mu\left(\phi(x), \phi(y) \right) \theta\left( \mu(x,y)^{-1} \right) = \left(\chi(x)^{-1} \right)^{\phi(y)} \chi(y)^{-1} \chi(xy).
	\label{eq:mu} 
	\end{align}
\end{widetext}
	The proof can be found in Ref.~\cite{Passi2010}, but let us comment briefly on the quantities involved: $\chi(x)$ is a map from $G$ to translations modulo $M$ ($V$ denotes the vector space of translations defined in the usual sense), it will depend in general on the choice made for the coset representative $t(x)$ of the point group elements $x \in G$, but once this choice is made $\chi(x)$ is unique and it is defined as $\gamma((t(x),e)) = (\phi(x),\chi(x))$. 
	Clearly, $\chi$ will also be acted on by $\theta$, but this action can be basically recast as a different choice for $t(x)$, hence we omit this dependence in the equations above.
	In practice, $\chi(x)$ can either be chosen to be identically zero if the space group is symmorphic and the Bravais lattice is primitive, or it is a known function otherwise.
	
	\autoref{eq:compAction} is a restatement of the compatible pair condition, whereas \autoref{eq:mu} provides a transformation law for the \mbox{2-cocycles} under the action of a compatible pair: the left hand side of \autoref{eq:mu} is a \mbox{2-cocycle} (let's call it $\mu^{(\theta,\phi)}(x,y)$). 
	Then, the condition $\gamma \in \ker \{ \tau \}$ corresponds to the case $\mu^{(\theta,\phi)}(x,y) = e$, \ie the extension $G^*$ is a direct product of the stabiliser of the wavevector with a group of translations, whereas for $\gamma \in \rm{im} \{ \tau\}$ one has $\mu^{(\theta,\phi)}(x,y) \neq e$ and the calculation of the little group of the wavevector can proceed in analogy with Herring's method.
	
	When $\gamma \in \ker \{\tau \}$, \autoref{eq:mu} is identically zero and a generator $t$ for the translation group is needed; 
	to this end we use Hopf's formula as reported in Ref.~\cite{Eick2008}.
	The element $t$ will then belong to $(G^*)'\cap M$, where $(G^*)'$ is the commutator subgroup of $G^*$. 
	If the extension is Abelian, $(G^*)'$ is trivial and no such translation exists.
	In this case the local symmetry group of $\kk$ coincides with its stabiliser.
	
	On the other hand, when $t \neq e$, we are left with the task of determining the order of such a translation group, that is the integer $p$ such that $t^p = e$.
	While in the Herring's method such a choice is made (heuristically) as summarised in \autoref{ssec:ThSum}, for the case of ``non-universal" $\kk$-points we make the estimate  for $|G^*| = p|G|$, with $p$ the smallest prime factor of the point group order $|G|$, thus removing the lattice parameters' dependence in the $\kk$-point coordinates: we denote this ``abridged" wavevector $\tilde{\kk}$.
	In this way, we are separating out the effect of the non-trivial translations on the stabiliser of the $\kk$-point from the trivial translations that depend on the specifics of the lattice parameters.

	This choice is motivated by the fact that an upper bound on the order of $G^*$ is given by the order of the semidirect product $M \rtimes G$, which is just the product of the orders of the two groups $M$ and $G$.
	The constraint on $|M|$ in this case originates from the requirement that $|M|$ and $|G|$ ought not to be coprime: if that were the case the resulting second cohomology group $H^2(G,M)$ would be trivial and the group extension $G^*$ would split.
	On the other hand, with our choice we ensure that a more complex structure of the extension group could be captured (since $H^2(G,M)$ is not trivial) while keeping the order of the resulting group minimal.
	In \autoref{sec:results} we will provide an example when the minimal choice for the group order is too restrictive and values of $p^n$ with $n > 1$ natural, have to be considered instead.
	
	Our reasoning is, perhaps, better explained with an example: let us consider the X-point for the Orthorhombic body-centred lattice, with coordinates $[- \xi	, \xi, \xi]$ (see \autoref{fig:oS}).
	By setting the lattice parameters to $a = \frac{1}{5}, \; b = \frac{1}{4} \; c = \frac{1}{3}$, one gets $\xi = \frac{17}{50}$, hence the phase factor $s = \exp [2\ii \pi \kk \cdot m]$ will be equal to 1 only for multiples of 50. 
	A translation group of order 50 contains a subgroup of order 2 and a subgroup of order 25 (the order of a subgroup must divide the order of the group), with the latter acting trivially on $G$, because $G$ can not contain five-fold rotations (owing to the crystallographic restriction) and thus its order $|G|$ will be coprime to $\textrm{5}^\textrm{2}$.
	The translation subgroup of order 2, on the other hand, can act non-trivially on $G$ and the resulting extension $G^*$ is what we tabulate. 
	In order for the $s$ coefficients to be able to reflect the periodicity of the translation subgroup acting non-trivially, we evaluate them using the ``abridged" wavevector $\tilde{\kk}=\frac{1}{p}[1,1,1]$, for the example at hand, and where $p=\mathrm{2}$ in this example, has been introduced previously.
	In general, the construction of the abridged vector replaces only the lattice dependent coordinates with the factor $\frac{1}{p}$.
	 
	To give an overall summary of the approach we employ (which is discussed in refs. \cite{Passi2010, HCGT}) we can then say that in order to construct an extension $G^*$ the knowledge of the \mbox{2-cocycles} $\mu(x,y)$ is required owing to \autoref{eq:prd}: to get a handle on these translations it is useful to study the behaviour of the resulting extension (in \autoref{eq:ExtSeq}) under automorphisms. 
	A special class of such automorphisms is $\mathsf{Comp}(G^*,M)$: the different (non-isomorphic) extensions $G^*$ correspond to the orbits of $\mathsf{Comp}(G^*,M)$ on $H^2_\alpha(G,M)$ (for a proof see $\S$ 2.7.4 in ref. \cite{HCGT}).
	The knowledge of $H^2_\alpha(G,M)$ is thus not necessary, as we only need to characterise the behaviour of $\gamma \in \Aut_M(G^*)$ with respect to it. 
	To this end, one employs the exact sequence in \autoref{eq:ExSeq2}, with the map $\tau$ playing an important role, since it allows to enumerate the two cases that can occur: if $\gamma \in \ker \{\tau \} = \mathrm{im}\{\psi\}$, then the automorphism $\gamma$ takes values in $Z_\alpha^1$, implying $\chi(xy) = \chi(x)^y \chi(y)$ and thus \autoref{eq:mu} is equal to the identity. 
	In this case the corresponding extension $G^*$ splits and we have suggested a scheme for the evaluation of such direct product in the previous paragraph.
	For the case when $\gamma \in\mathrm{im}\{\tau\}$ one has that the function $\chi(x)$ forms a \mbox{2-coboundary} and \autoref{eq:mu} provides already a recipe for the construction of the Schur's multipliers as the function $\chi(x)$ is a known function over the entire space-group.
	
\section{Computational Methods}
\label{sec:comp}

	In our code, we tabulate the generators of the 230 space groups in the conventional unit cell settings in accordance with the International Tables for Crystallography (vol. A) \cite{ITCA}.
	Consistently, we identify the $\kk$-point coordinates in the Wigner-Seitz reciprocal lattice unit cell of the additional wavevectors reported in Ref.~\cite{Setyawan2010}.
	For completeness we report in \autoref{fig:mP}-\ref{fig:cF} these Brillouin zones and the coordinates of the wavevectors with respect to the primitive basis vectors (also reproduced in the pictures).
	The algorithm then proceeds to construct the stabiliser of the wavevector and hence its little group, following either of the methods reported in \autoref{sec:Th}.
	
	In a limited number of instances, the local symmetry group generated can be bigger than our theoretical estimate outlined previously: this happens when the non-trivial translation $\mu^{(\theta,\phi)}$ is collinear with the centering vectors in the non-primitive unit cell. 
	For non-symmorphic space groups, the action of the centering translations can be non trivial and the resulting order of the extension exceeds the estimate by a factor proportional to the number of centering vectors present in the resulting group, typically by a factor two, thus leading to a local symmetry group of order $p^2|G|$.
	In order to better investigate the group structure (see the discussion in \autoref{sec:results}) the algorithm can list all subgroups of index $n$ thanks to a one-to-one correspondence with (standardised and complete) coset tables with $n$ rows and the subgroups of the given group; for a detailed exposition of the backtrack search strategy employed in the algorithm's implementation we refer to chapter 5 in Refs.~\cite{HCGT,Sims1994} or to Ref.~\cite{Cannon1973a}.
	
\begin{figure*}
	\includegraphics[width=140mm]{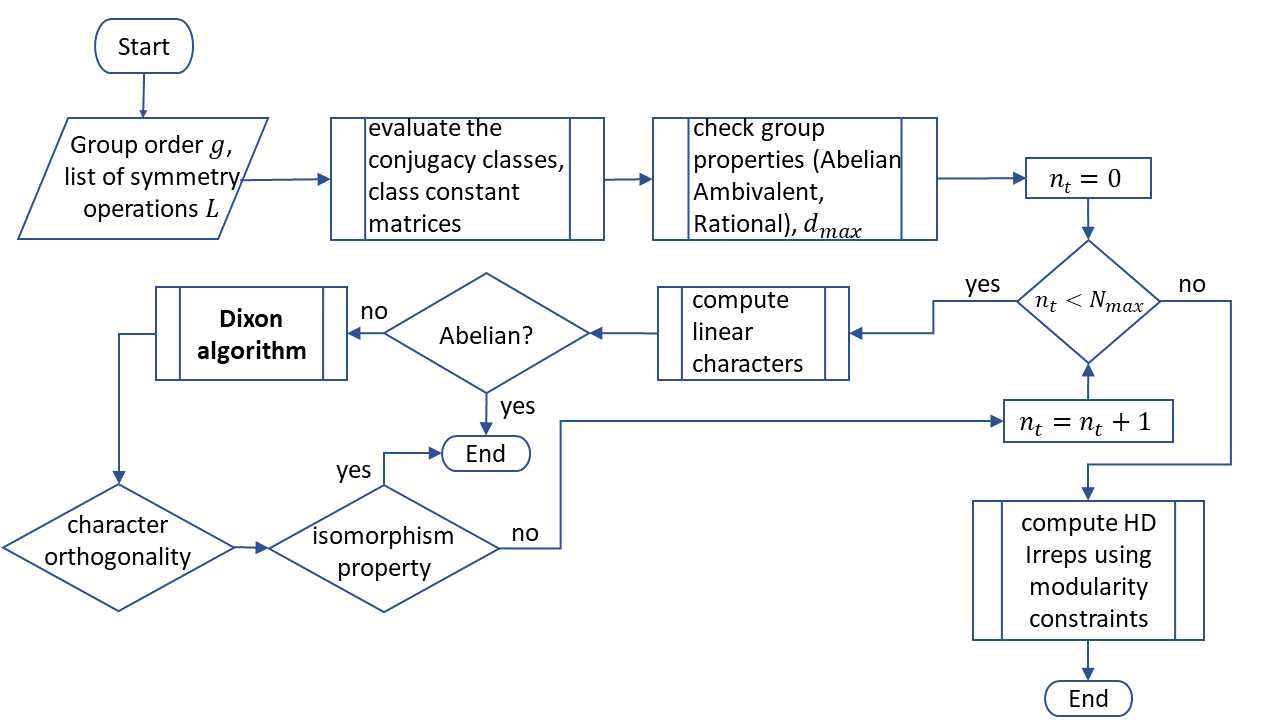}
	\caption{Flow chart for the algorithm computing the character tables. A preliminary set of operations is carried out at the beginning of the code, which include the evaluation of conjugacy classes and maximum allowed dimension for irreducible representations (Irreps) $d_{max}$. If the group is not Abelian, the Dixon algorithm tries to compute the higher dimensional (HD) Irreps, \ie Irreps with degree bigger than 1. If the number of iterations $n_t$ exceeds a threshold ($N_{max}$), the program attempts to compute HD characters using modularity constraints in the search.}
	\label{fig:Dixon}
\end{figure*}

	Once that the little group of the wavevector has been constructed we determine its irreducible representations (over the complex numbers) by computing its character table.
	To this end we have implemented the Dixon algorithm \cite{Dixon1967,Dixon1970,HCGT,Grove1997}, which makes use of modular arithmetic to efficiently diagonalise the class constant matrices.
	 In order to simultaneously diagonalise the class constant matrices, the order in which the individual matrices are fed into the algorithm plays a role, since it is possible for a given matrix to have eigenvalues with multiplicity bigger than one and such degeneracy can be resolved by prompting only a specific choice for the next matrix (see $\S$ 7.7 in ref.  \cite{HCGT}).
	 Furthermore, the expression for the class matrices themselves depends on the choice for the group's generators \cite{El-Batanouny2008}. 
	 Selection at random for the generating elements and for the permutation order of the matrices typically allows to retrieve the full character table within a few attempts; should that not be the case, we proceed by computing the characters of the group Abelianisation and then enforce congruences among characters of high dimensional irreducible representations still missing, as suggested in Ref.~\cite{Lux2010}. 
	 In \autoref{fig:Dixon} we report a flow chart for our implementation.
	 
	We cross-check our implementation against the tabulated results for universal $\kk$-points in Ref.~\cite{Bradley1972a}. 
	Additionally, we also check that the character table preserves the group multiplication rule, that is the  computed characters are actually an homomorphism $\zeta : G^* \to \mathbb{C} $.
	This is enforced by requiring the convolution of characters to fulfill the orthogonality condition \cite{Curtis1962}:
	
	\begin{align}
	 \sum_{g\in G^*} \overline{\zeta_r}(g) \zeta_s(hg) = \frac{\delta_{rs} \zeta_r(h)|G^*|}{\zeta_r(e)} 
	 \label{eq:chiConv} 
	\end{align}
	for each element $h \in G^*$ and for all irreducible representations $r,s$, overbar denotes complex conjugation in the expression above. 

	Our algorithm automatically checks for \autoref{eq:chiConv} when generating the character table, besides the usual orthogonality relations, which are special cases of the equation above.
	We point out that \autoref{eq:chiConv} involves to globally verify the group structure, and it can be tested only laboriously taking as input the  character tables provided in Ref.~\cite{Bradley1972a} since  the corresponding abstract groups are tabulated therein.

	Finally we touch on the computational methods to obtain the actual representation matrices starting from the character table.
	The algorithm that we are about to discuss is largely mutuated from the articles by Blokker \cite{Blokker1972,Blokker1973} and we also refer to the classic ref. \cite{Curtis1962} for an in-depth discussion of the theoretical aspects.
	Denoting $\Gamma^r$ the regular representation of a group $G$ one can form the sum of elements in the $i^\mathrm{th}$ conjugacy class, which we call $\mathfrak{C}_i$. 
	Then the operators $\mathfrak{p}_j$ are uniquely determined idempotent operators defined in the group algebra $\mathbb{C}G$:
	\begin{align}
	\mathfrak{p}_j = \frac{d_j}{|G|} \sum_{i=1}^n \overline{\zeta_j}(g_i) \mathfrak{C}_i
	\label{eq:proj}
	\end{align}
	where $n$ is the number of conjugacy classes, $d_j=\zeta_j(e)$ is the representation degree and $g_i$ is a representative element for the  $i^\mathrm{th}$ class. 
	The set of class functions is spanned by the centre of the group algebra $Z[\mathbb{C}G]$, hence our objective is to form a representation for $Z[\mathbb{C}G]$ by projecting the regular representation over the basis $\mathbf{t}_i$ for the range of the operator $\mathfrak{p}_i$:
	\begin{align*}
	\Gamma^Z_i(g) = \mathbf{t}_i^\dagger \Gamma^r(g) \mathbf{t}_i
	\end{align*}
	the matrices  $\Gamma^Z_i$ will in general be reducible representations of degree $d_i^2$ containing the irreducible representation $\Gamma_i$ only $d_i$ times. The associated eigenvalues will then have multiplicity $d_i$
     and the corresponding eigenvectors  form an orthonormal basis $\mathbf{B}_g^{(i)}$ for the subspace of $Z[\mathbb{C}G]$ associated with the group element $g \in G$. 
     Since the projection operators commute with the group elements, to obtain the other entries of the basis one can act with the remaining elements of $G$ to span the whole centre of the group algebra. 
	Finally, the irreducible representations matrices are obtained by a similarity transform:
	\begin{align}
	\Gamma_i(g) = (B_g^{(i)})^\dagger \Gamma_i^Z(g) B_g^{(i)}.
	\label{eq:Irreps}
	\end{align}
	
	In order to induce the space group representations, a further constraint has to be imposed on the irreducible representations obtained in \autoref{eq:Irreps}: $\zeta(\mu) \neq \zeta(e)$, that is, the lattice translation $\mu$ must not act trivially by having its character  belonging to the kernel of the irreducible representation.	
	All matrix representations for the local symmetry groups computed in this work are reported in the supporting information, and the selection of the allowed representations is left to the final user.

\section{Results and discussion}
\label{sec:results}

	We identify the (non-trivial) little group of the wavevectors corresponding to those first obtained in Ref.~\cite{Setyawan2010} for all space groups having the relevant Bravais lattice.

\begin{figure}
	\includegraphics[width=80mm]{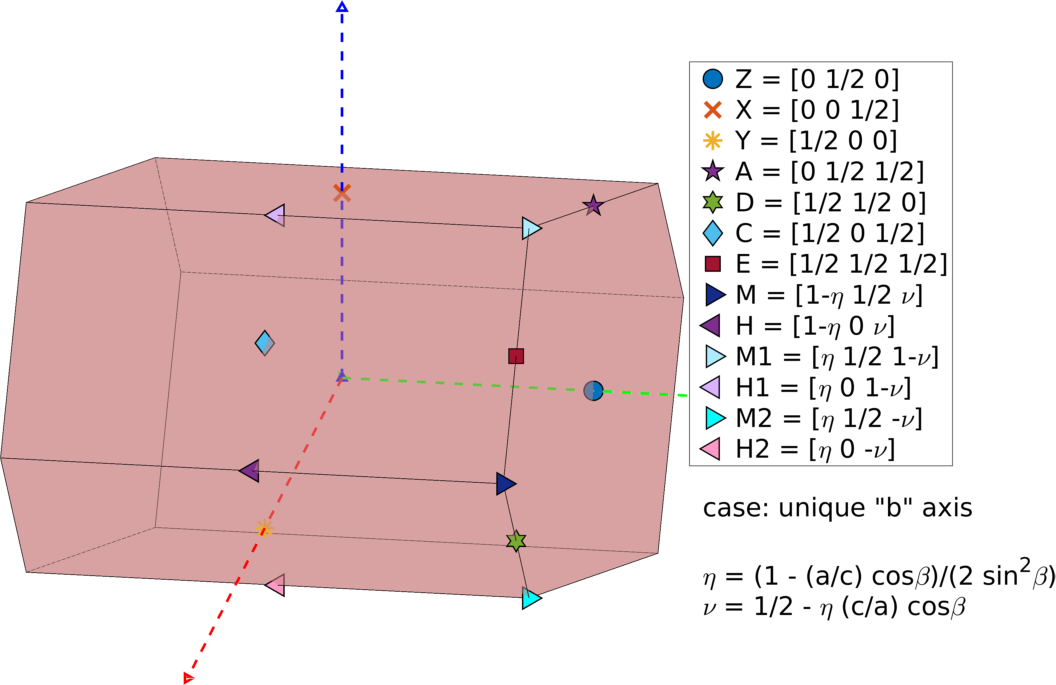}
	\caption{Brillouin zone and high symmetry points for the Monoclinic primitive lattice. Symmetry operations defined according to the 'unique b axis' crystallographic convention. Little groups of the wavevectors reported in \autoref{tab:m}.}
	\label{fig:mP}
\end{figure}

\begin{figure*}
	\includegraphics[width=120mm]{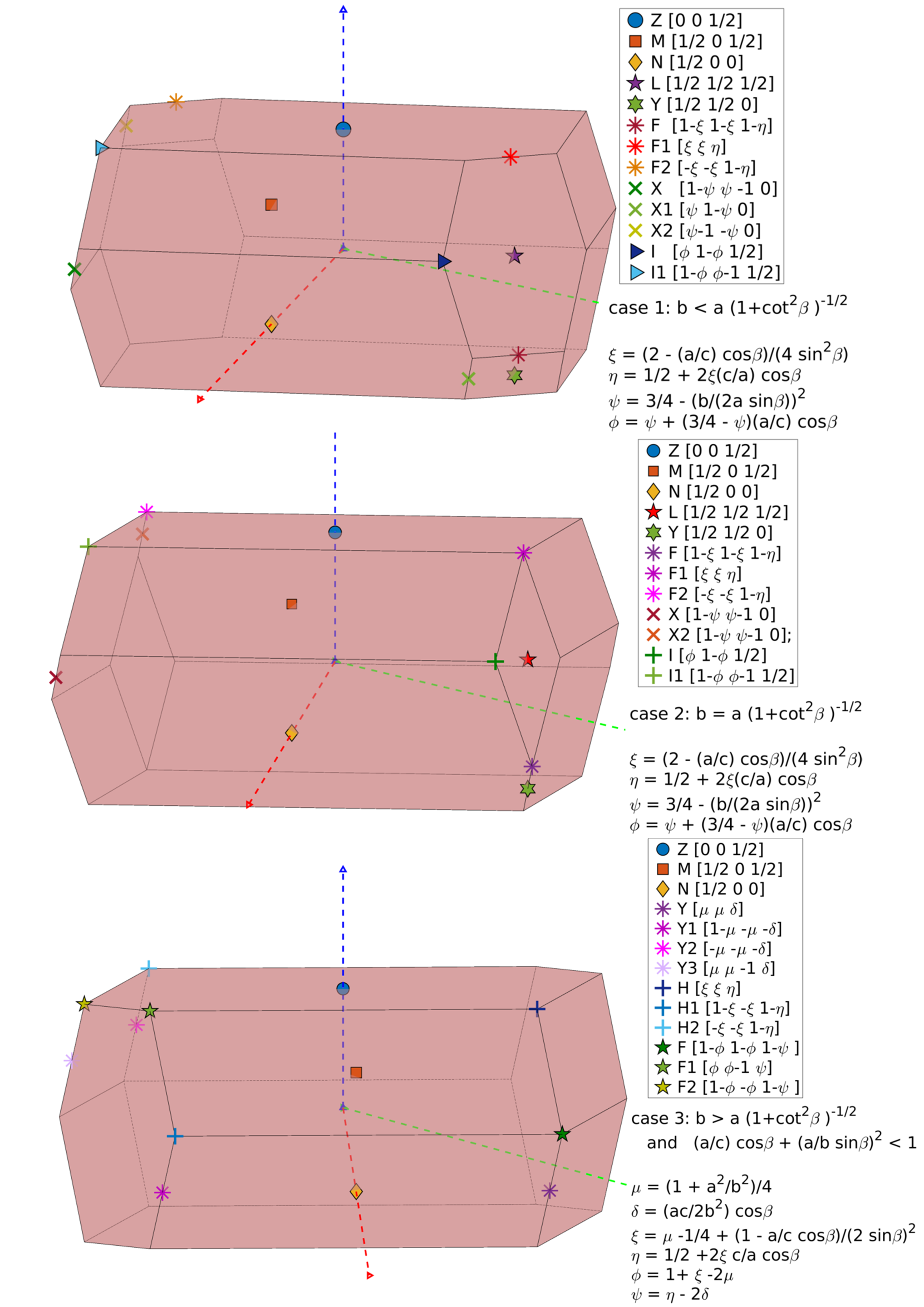}
	\caption{Brillouin zone and high symmetry points for the Monoclinic base-centred lattice. Cases 1-3 of Ref.~\cite{Setyawan2010} shown from top to bottom. Little groups of the wavevectors reported in \autoref{tab:m}.}
	\label{fig:mC1-3}
\end{figure*}

\begin{figure*}
	\includegraphics[width=120mm]{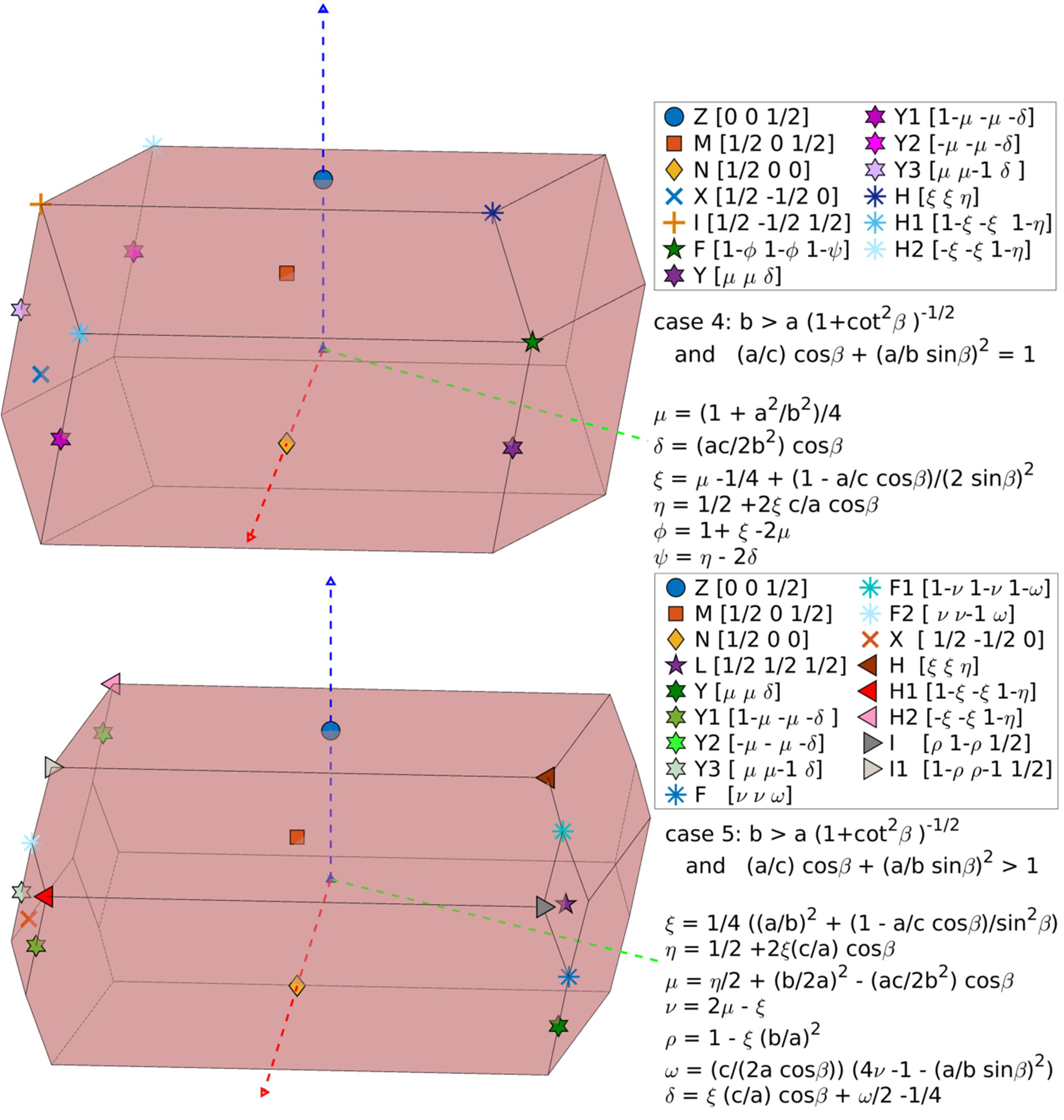}
	\caption{Brillouin zone and high symmetry points for the Monoclinic base-centred lattice. Cases 4-5 of Ref.~\cite{Setyawan2010} shown from top to bottom. Little groups of the wavevectors reported in \autoref{tab:m}.}
	\label{fig:mC4-5}
\end{figure*}

\begin{figure}
	\includegraphics[width=75mm]{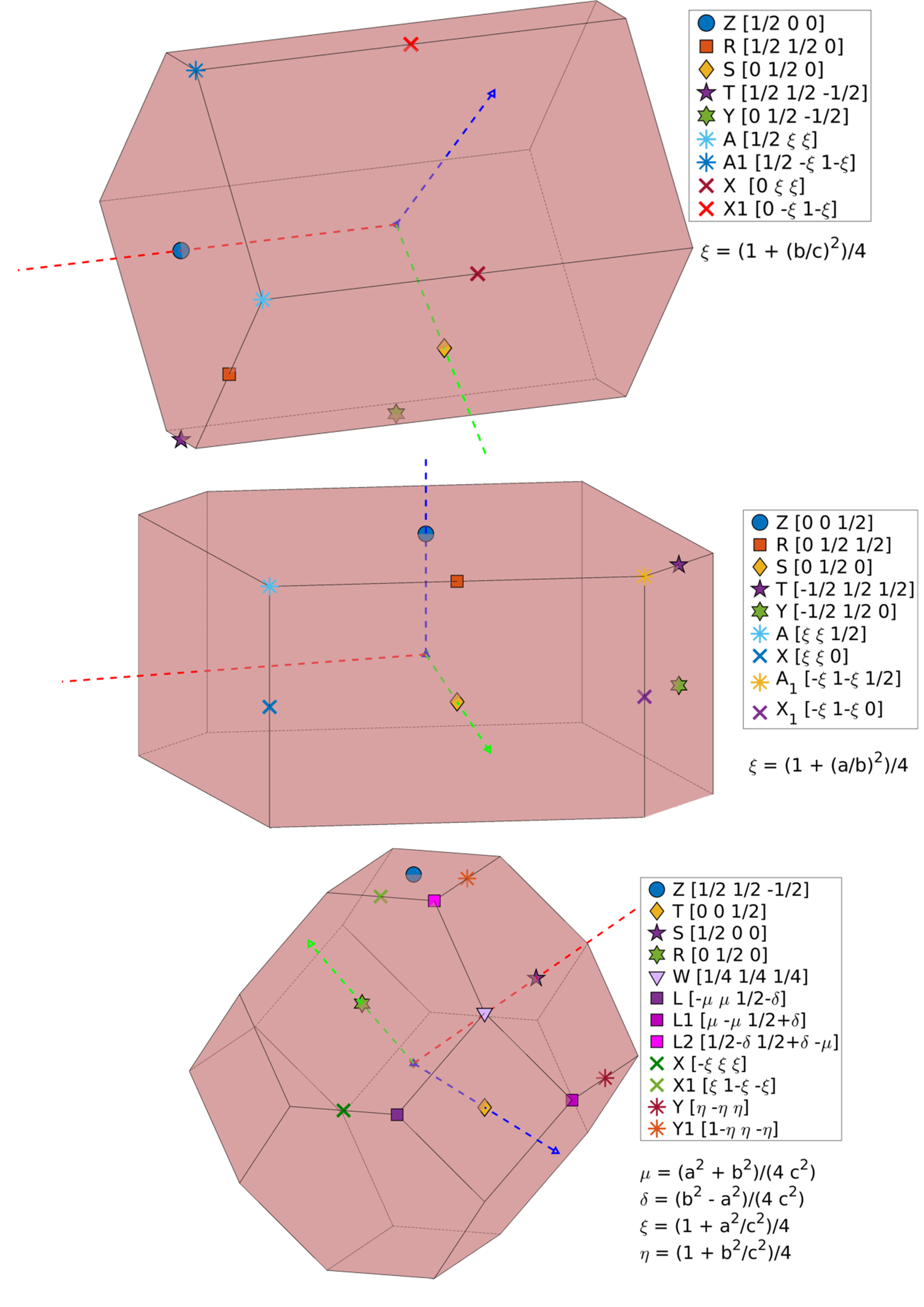}
	\caption{Brillouin zone and high symmetry points for the Orthorhombic oA lattice (top), oC lattice (middle), oI lattice (bottom). Little groups of the wavevectors reported in \autoref{tab:oS}.}
	\label{fig:oS}
\end{figure}

\begin{figure}
	\includegraphics[width=75mm]{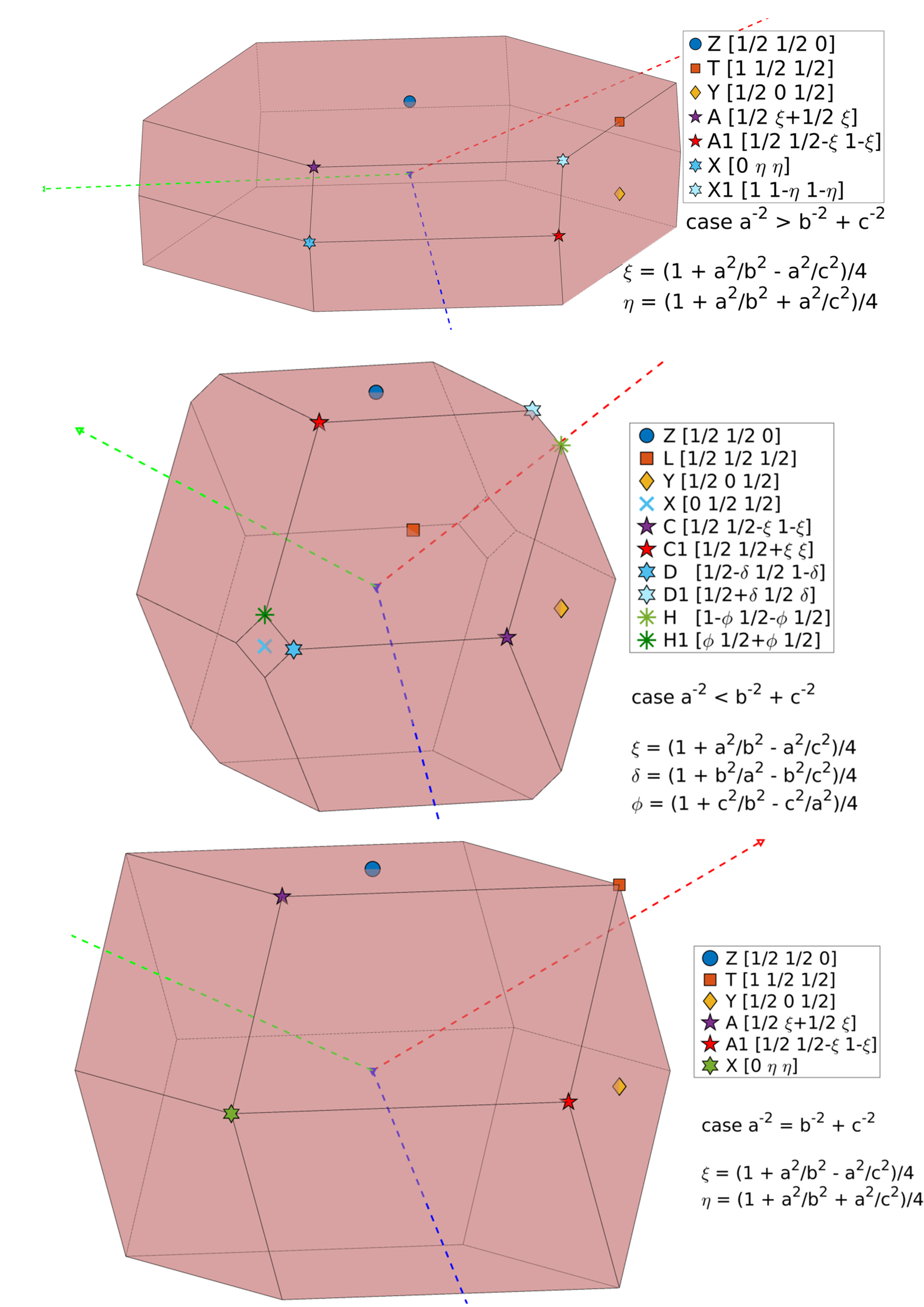}
	\caption{Brillouin zone and high symmetry points for the Orthorhombic face-centred lattice.  Cases 1-3 of Ref.~\cite{Setyawan2010} shown from top to bottom. Little groups of the wavevectors reported in \autoref{tab:oF}.}
	\label{fig:oF}
\end{figure}

\begin{figure}
	\includegraphics[width=75mm]{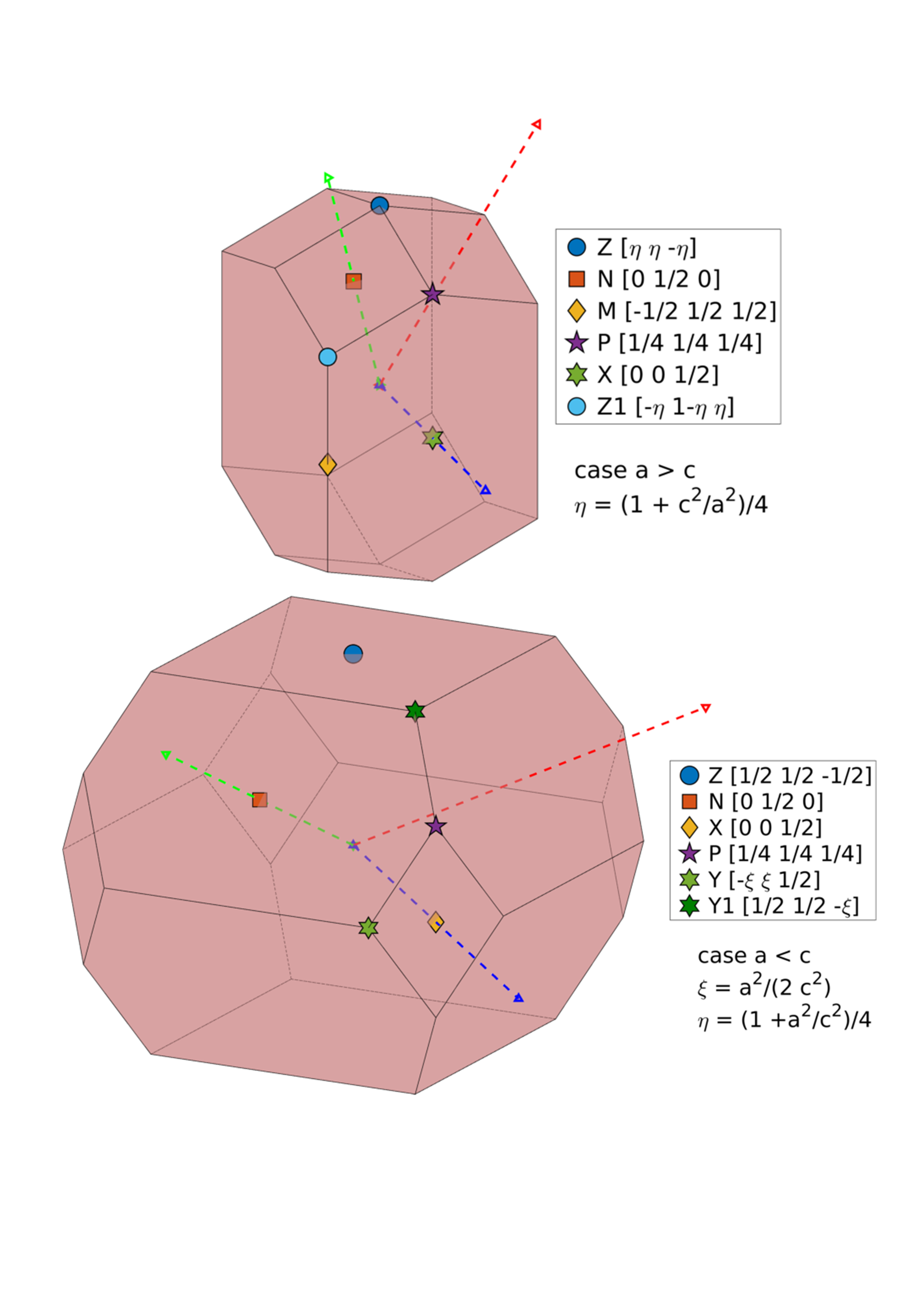}
	\caption{Brillouin zone and high symmetry points for the Tetragonal body-centred lattice. Cases 1-2 of Ref.~\cite{Setyawan2010} shown from top to bottom. Little groups of the wavevectors reported in \autoref{tab:tI}.}
	\label{fig:tI}
\end{figure}

\begin{figure}
	\includegraphics[width=70mm]{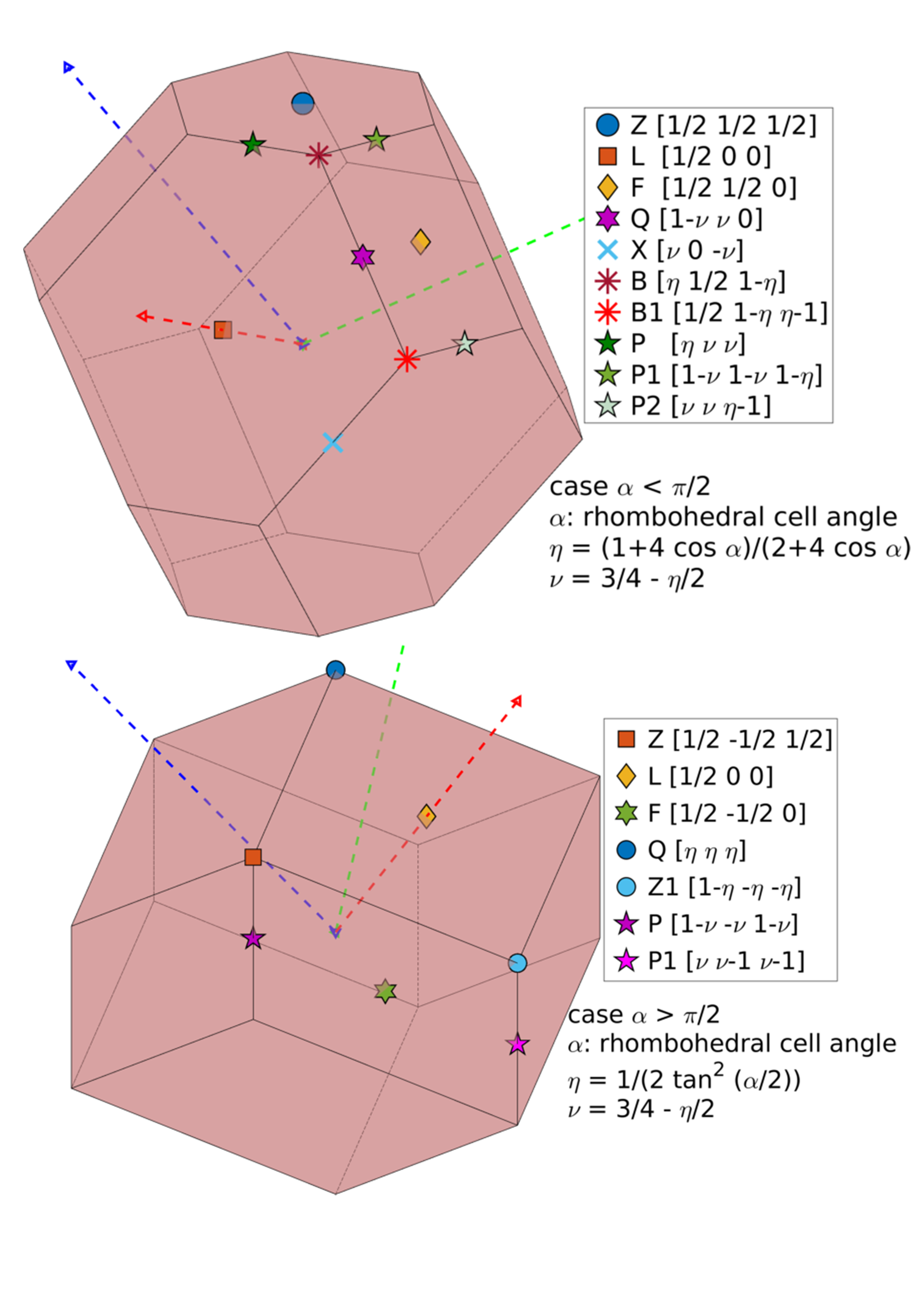}
	\caption{Brillouin zone and high symmetry points for the Hexagonal rhombohedral lattice. Cases 1-2 of Ref.~\cite{Setyawan2010} shown from top to bottom. Little groups of the wavevectors reported in \autoref{tab:hR}.}
	\label{fig:hR}
\end{figure}

\begin{figure}
	\includegraphics[width=80mm]{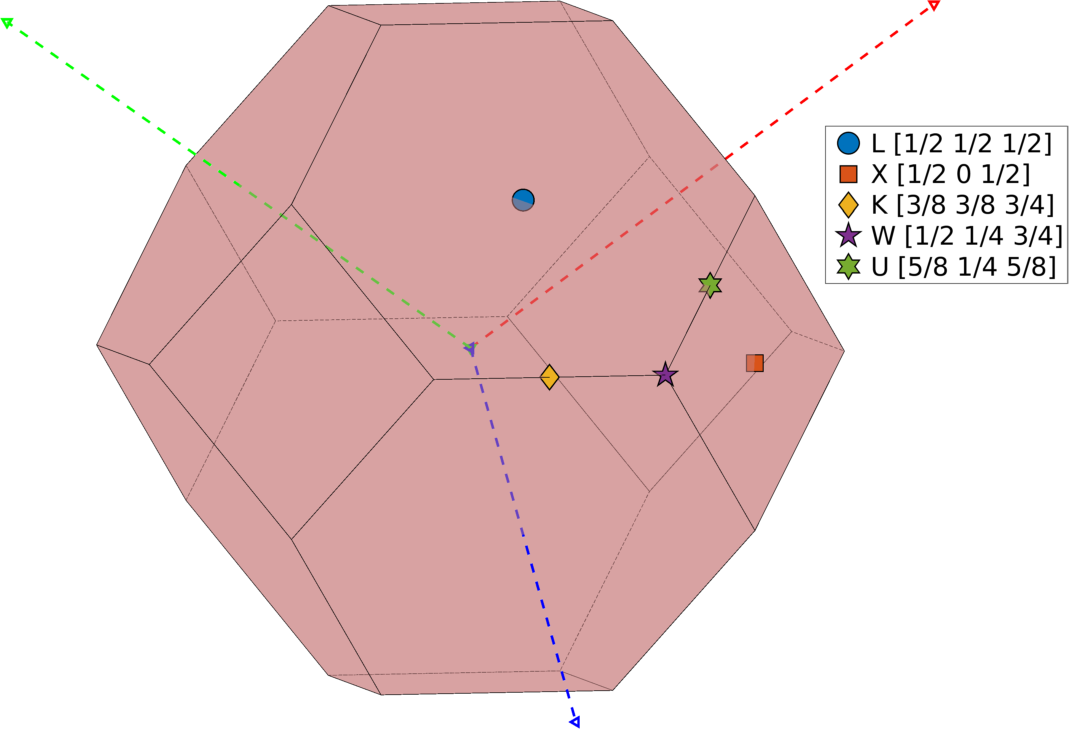}
	\caption{Brillouin zone and high symmetry points for the Cubic face-centred lattice. Little groups of the wavevectors reported in \autoref{tab:cF}.}
	\label{fig:cF}
\end{figure}

	These groups are listed in tables \ref{tab:m} to \ref{tab:cF}, for completeness we also report the corresponding Brillouin zones and wavevectors coordinates in Figs. \ref{fig:mP} to \ref{fig:cF}, as we follow the standard crystallographic convention of the International Tables of Crystallography.
	In particular, for the orthorhombic lattice we consider only the so-called standard setting with $a < b < c$ out of the six settings that are allowed by symmetry in orthorhombic systems \cite{ITCA}.
	Any non-standard choice for the lattice vectors orientation simply rotates the Brillouin zone in reciprocal space, leaving its shape unaffected, and so are the local symmetry groups at the ``non-universal'' wavevectors.
		
	For each of the space groups considered we report the generators of the Abelian little groups (tables \ref{tab:Gen_m}-\ref{tab:Gen_cF}, the notation $\mathbf{0}$ therein indicates the null translation) and character tables for wavevectors that have a non-Abelian little group (tables \ref{table:SG15X}-\ref{table:SG228K}, the notation $\zeta_n$ therein indicates the $n$-th root of unity): 
	we decided to individually tabulate the character table for each of these groups, even though two (or more) of them might be isomorphic. 
	This is to make explicit the connection between the classes of the abstract group and the symmetry operation of the specific little group at hand.

	In the previous section we have mentioned the instance of our algorithm identifying a local symmetry group of order exceeding the simple estimate provided in \autoref{sec:Th}, which occurs when the inclusion of centering vectors (that belong to the same coset as the identity element once the lattice translations have been factored out the space group) is necessary when non-symmorphic symmetry elements are present.
	As an example, consider the A wavevector for space group \#63: if no restrictions are imposed on the local symmetry group order, the program identifies, starting from a point symmetry of order 4, a group of order 16, featuring two centering vectors $t_1=(\frac{1}{2}, \frac{1}{2},0)$ and $t_2=(\frac{1}{2}, \frac{1}{2},1)$.
	The translations $t_1$ and $t_2$ are non-equivalent, as it can be checked by evaluating the Herring phase factors, and there is no group operation that conjugates them, as such they must belong to singleton conjugacy classes; since the group does not contain any symmetry element of order 8, one can readily identify the group in question as the Pauli group $\ZZ_4 \circ \textsf{D}_\textsf{8}$.
	
	To further verify the correctness of the group structure presented above we compute the  presentation for the group $\ZZ_4 \circ \Dn$:  this consists of the generating elements: 
	\begin{enumerate}
	\item$ (xyz|t_2)$, 
	\item$ (xyz|110)$, 
	\item$ (x \yb \zb | 000)$, 
	\item$ (x y \zb | 0 0 \frac{1}{2})$, 
	\end{enumerate} 
	together with a set of relators (computed following ref. \cite{Cannon1973}) that enforce constraints on the group structure.
	Among these relators, particularly crucial is the relationship 434 = 32 (where the numbers refer to the generators as listed above and the group product is the Seitz rule), thanks to this relator one can identify the two translations $(xyz | 001)$ and $(xyz | 110) = t_H$: this is a particular instance of the fact that the Herring mapping for the construction of the extension $G^*$ is not necessarily consistent with the Seitz rule in \autoref{eq:prd} being taken as the group operation, as pointed out in Refs.~\cite{Jones1973,Koster1957}.	
	This happens since for an entry $[x,s]$ there might be more than one translation $m \in M$ that are identified by the same value of the Herring phase factor $s$.
	
	The quaternion group $\Qt$ can hence be constructed from the two elements $c = (xy \zb | \frac{1}{2} \frac{1}{2} \frac{1}{2})$ and $b = (x \yb z | 0 0 \frac{1}{2})$; 
	in particular one can verify that $c^4 = e, b^2 = c^2 = t_H$  and that the action of one element on the other is non-trivial with $bcb^{-1} = c^{-1}$. 
	Another subgroup that can be readily identified is $\Dn$, by choosing the elements $a = (xy \zb | 0 0 \frac{1}{2})$ and $b$ as above.
	By direct calculation one can observe that $a^2 = b^4 = e$ and $aba = b^{-1}$, as long as the Herring translation is now taken to be $t_H = (xyz | 0 0 1)$.
	The overall symmetry group $\ZZ_4 \circ \Dn$ contains both groups $\Qt$ and $\Dn$ as normal subgroups of index two and only by considering this larger symmetry group one has a complete picture of the symmetries at play.
	 For space group \#63, these can thus be rationalised as arising from the presence of two equivalent translation vectors (along the $z$-direction and along the $x+y$-direction) unrelated by the space group operations.
	 
	During the submission stage, we became aware of a recent publication \cite{Zhang2023} that addresses a related problem of evaluating the action of non-symmorphic symmetry operations on generic wavevectors with the aid of projective representations. 
	Since symmetries in direct space (considered in this work) and in reciprocal space (considered in ref. \cite{Zhang2023}) can be put in correspondence with one another, projective representations and the identification of the appropriate extension with the Herring method are concurrent for the determination of the local symmetry group's irreducible representations.
	
	In conclusion, we have proposed an algorithm for the evaluation of the local symmetry group for arbitrary $\kk$-points, including those whose coordinates explicitly depend on lattice parameters and that for that reason could not be dealt with using the consolidated approach by Herring.
	We think that our code, that we are planning to release to the general public, could be a useful development to be integrated in the induction of space group representations and can help strengthen the connection between symmetry indicators and their evaluation as Berry phases along closed loops in the Brillouin zone \cite{Song2018a}, along which ``non-universal" $\kk$-points can be found.

\section*{Acknowledgments}
	The authors acknowledge support from the Austrian Science Fund (FWF) through project BandITT P 33571.
Calculations were performed in part on the Vienna Scientific Cluster (VSC).
The authors acknowledge TU Wien Bibliothek for financial support through its Open Access Funding Programme.
  
\bibliography{Books,GroupTheory,Topology}
  

\begin{table*}[t]
\begin{center}
\begin{minipage}{\textwidth}
\caption{``non-universal" wavevectors' little groups for space groups with Monoclinic Bravais lattices. Plots of the corresponding Brillouin zones are reported in Fig.\ref{fig:mP} (primitive) and Figs.\ref{fig:mC1-3}-\ref{fig:mC4-5} (base-centred), together with the relations defining the respective cases.} \label{tab:m}
\begin{tabular*}{\textwidth}{@{\extracolsep{\fill}}rlllcc@{\extracolsep{\fill}}}
\toprule
\multicolumn{2}{@{}c@{}}{Space group} & Cases 1, 2 & Cases 3, 4 & Case 5 \\
\midrule
5 	& 	C2	 &$\textrm{\{X, I \}} \cong \ZZ_2$ & $\textrm{\{F \}} \cong \ZZ_2$ & $\textrm{\{I \}} \cong \ZZ_2$ \\

6 	& 	Pm 	 &$\textrm{\{M, M1, H, H1 \}} \cong \ZZ_2$ & & \\

7  	&	Pc   &$\textrm{\{M, H, M1, H1 \}} \cong \ZZ_4$  & & \\

8 	&	Cm 	 &$\textrm{\{F, F1, X \}} \cong \ZZ_2$  &$\textrm{\{Y, F, H \}} \cong \ZZ_2$ & $\textrm{\{Y, F, H \}} \cong \ZZ_2$ \\

9	& 	Cc	 & $\textrm{\{F, F1, X \}} \cong \ZZ_4 $ & $\textrm{\{Y, F, H \}} \cong \ZZ_4 $ & $\textrm{\{Y, F, H \}} \cong \ZZ_2$ \\

10 	& 	P2/m &$\textrm{\{M, H, M1, H1 \}} \cong \ZZ_2$  & & \\

11 	& 	P2$_\textrm{1}$/m &$\textrm{\{M, H, M1, H1 \}} \cong \ZZ_2$  & & \\

12 	& 	C2/m & $\textrm{\{F, F1, I \}} \cong \ZZ_2 \; \textrm{ \{X  \} } \cong \ZZ_2 \x \ZZ_2$ & $\textrm{ \{Y, H \}} \cong \ZZ_2 \; \textrm{ \{F  \} } \cong \ZZ_2 \x \ZZ_2$ & $\textrm{ \{Y, F, H, I \}} \cong \ZZ_2 $ \\

13 	&	P2/c &$\textrm{\{M, H, M1, H1 \}} \cong \ZZ_4$  & & \\

14 	&	P2$_\textrm{1}$/c &$\textrm{\{M, H, M1, H1 \}} \cong \ZZ_4$  & & \\

15& 	C2/c& $\textrm{\{F, F1, I \}} \cong \ZZ_4 \; \textrm{\{X \}} \cong \ZZ_2 \x \Dn$  & $\textrm{\{Y \}} \cong \ZZ_4 \; \textrm{\{F \}} \cong \ZZ_2 \x \Dn \; \textrm{\{H \}} \cong \ZZ_2 $ & $\textrm{\{Y, F, H, I \}} \cong \ZZ_4$ \\

\bottomrule
\end{tabular*}
\end{minipage}
\end{center}
\end{table*}

\begin{table*}[p]
\begin{center}
\begin{minipage}{\textwidth}
\caption{``non-universal" wavevectors' little groups for space groups with Orthorhombic base-centred and body-centred Bravais lattices. Plots of the corresponding Brillouin zones are reported in Fig.\ref{fig:oS}. } \label{tab:oS}
\begin{tabular*}{\textwidth}{@{\extracolsep{\fill}}rlllcc@{\extracolsep{\fill}}}
\toprule
\multicolumn{2}{@{}c@{}}{Space group} & Case 1 \\
\midrule
20 	&	C222$_\textrm{1}$	&$\textrm{\{A, A1, X, X1 \}} \cong \ZZ_4 $ \\

21 	& 	C222 & $\textrm{\{A, A1, X, X1 \}} \cong \ZZ_2 $ \\

23	&	I222 & $\textrm{\{X, Y \}} \cong \ZZ_2$ \\

24 	& 	I2$_\textrm{1}$2$_\textrm{1}$2$_\textrm{1}$ & $ \textrm{\{X,Y \}} \cong \ZZ_4$ \\

35	& 	Cmm2 & $\textrm{\{A, A1, X, X1 \}} \cong \ZZ_2$ \\

36	& 	Cmc2$_\textrm{1}$ & $\textrm{\{A, A1, X, X1 \}} \cong \ZZ_4 $  \\

37  & 	Ccc2 & $\textrm{\{A, A1, X, X1 \}} \cong \ZZ_4 $ \\

38	&	Amm2 &	$\textrm{\{A, X \}} \cong \ZZ_2$ \\

39  & 	Aem2 & $\textrm{\{A, X \}} \cong \ZZ_4 $\\

40  & 	Ama2 & $\textrm{\{A, X \}} \cong \ZZ_4 $\\

41  & 	Aea2 & $\textrm{\{A, X \}} \cong \ZZ_4 $\\

44  & 	Imm2 & $\textrm{\{X, Y \}} \cong \ZZ_2 $\\

45  & 	Iba2 & $\textrm{\{X, Y \}} \cong \ZZ_4 $ \\

46  & 	Ima2 & $\textrm{\{X, Y \}} \cong \ZZ_4 $\\

63  & 	Cmcm & $\textrm{\{A \}} \cong \ZZ_4 \circ \textsf{D}_\textsf{8} \; \textrm{\{X \}} \cong \ZZ_2 \x \ZZ_2 \x \ZZ_4 \; \textrm{\{A1, X1 \}} \cong \ZZ_2 \x \Dn $ \\

64  & 	Cmce & $\textrm{\{A, A1, X1 \}} \cong \ZZ_4 \circ \Dn \; \textrm{\{X \}} \cong \ZZ_2 \x \ZZ_2 \x \ZZ_4 $ \\

65 	& 	Cmmm & $\textrm{\{A, X, A1, X1 \}} \cong \ZZ_2 \x \ZZ_2 $ \\

66  & 	Cccm & $\textrm{\{A \}} \cong \ZZ_4 \circ \textsf{D}_\textsf{8} \; \textrm{\{X \}} \cong \ZZ_2 \x \ZZ_2 \x \ZZ_4 \; \textrm{\{A1, X1 \}} \cong \ZZ_2 \x \Dn $ \\

67  & 	Cmme & $\textrm{\{A, X \}} \cong \ZZ_2 \x \ZZ_2 \x \ZZ_4 \; \textrm{\{A1, X1 \}} \cong \ZZ_4 \circ \Dn $ \\

68  & 	Ccce & $\textrm{\{A, A1, X1 \}} \cong \ZZ_4 \circ \Dn \; \textrm{\{X \}} \cong \ZZ_2 \x \ZZ_2 \x \ZZ_4 $ \\

71	& 	Immm & $\textrm{\{L \}} \cong \ZZ_2 \; \textrm{\{X, Y \}} \cong \ZZ_2 \x \ZZ_2 $ \\

72  & 	Ibam & $\textrm{\{L \}} \cong \ZZ_2 \; \textrm{\{X, Y \}} \cong \ZZ_2 \x \ZZ_2 \x \ZZ_4 \; \textrm{\{X1, Y1 \}} \cong \ZZ_4 \circ \Dn $ \\

73  & 	Ibca & $\textrm{\{L, L1, L2 \}} \cong \ZZ_4 \; \textrm{\{X, Y \}} \cong \ZZ_2 \x \ZZ_2 \x \ZZ_4 \; \textrm{\{X1, Y1 \}} \cong \ZZ_4 \circ \Dn$ \\

74  & 	Imma & $\textrm{\{L, L1, L2 \}} \cong \ZZ_4 \; \textrm{\{X, Y \}} \cong \ZZ_2 \x \ZZ_2 \x \ZZ_4 \; \textrm{\{X1, Y1\}} \cong \ZZ_4 \circ \Dn$\\

\bottomrule
\end{tabular*}
\end{minipage}
\end{center}
\end{table*}	 

\begin{table*}[p]
\begin{center}
\begin{minipage}{\textwidth}
\caption{``non-universal" wavevectors' little groups for space groups with Orthorhombic face-centred Bravais lattice. Plots of the corresponding Brillouin zones are reported in Fig.\ref{fig:oF}, together with the relations defining the respective cases. } \label{tab:oF}
\begin{tabular*}{\textwidth}{@{\extracolsep{\fill}}rlllcc@{\extracolsep{\fill}}}
\toprule
\multicolumn{2}{@{}c@{}}{Space group} & Cases 1,3 & Case 2 &\\
\midrule
22  & F222 & $\textrm{\{A \}} \cong \ZZ_2 \; \textrm{\{X \}} \cong \ZZ_2 \x \ZZ_2$	 & $\textrm{\{C, D, H \}} \cong \ZZ_2$ \\

42 	& Fmm2 & $\textrm{\{A \}} \cong \ZZ_2 \; \textrm{\{X \}} \cong \ZZ_2 \x \ZZ_2$ & $\textrm{\{C, D \}} \cong \ZZ_2 \; \textrm{\{H \}} \cong \ZZ_2 \x \ZZ_2$ \\

43  & Fdd2 & $\textrm{\{A \}} \cong \ZZ_2 \x \ZZ_4 \; \textrm{\{X \}} \cong \textsf{D}_\textsf{8} \rtimes \ZZ_4$ & $\textrm{\{C, D \}} \cong \ZZ_8 \; \textrm{\{H \}} \cong \ZZ_4 \wr \ZZ_2$\\

69  & Fmmm & 	$\textrm{\{A \}} \cong \ZZ_2 \x \ZZ_2 \; \textrm{\{X \}} \cong \ZZ_2 \x \ZZ_2 \x \ZZ_2 $ & $\textrm{\{C, D, H \}} \cong \ZZ_2 \x \ZZ_2$ \\

70  & Fddd & $\textrm{\{A \}} \cong \ZZ_2^3 \rtimes \ZZ_4 \; \textrm{\{X \}} \cong \ZZ_2^2  \rtimes \textsf{D}_\textsf{16}$ &	 $\textrm{\{C, D, H \}} \cong \ZZ_4 \wr \ZZ_2 $\\

\bottomrule
\end{tabular*}
\end{minipage}
\end{center}
\end{table*}	 

\begin{table*}[p]
\begin{center}
\begin{minipage}{\textwidth}
\caption{``non-universal" wavevectors' little groups for space groups with Tetragonal body-centred Bravais lattice. Plots of the corresponding Brillouin zones are reported in Fig.\ref{fig:tI}, together with the relations defining the respective cases. } \label{tab:tI}
\begin{tabular*}{\textwidth}{@{\extracolsep{\fill}}rlllcc@{\extracolsep{\fill}}}
\toprule
\multicolumn{2}{@{}c@{}}{Space group} & Case 1 & Case 2 &\\
\midrule
79  	&	I4 &	$\textrm{\{Z \}} \cong \ZZ_2 \x \ZZ_4$ & -\\

80  & I4$_\textrm{1}$ &  $\textrm{\{Z \}} \cong \ZZ_2 \x \ZZ_8$ & -\\

82  & I$\overline{\textrm{4}}$ & $\textrm{\{Z \}} \cong \ZZ_2$ & -\\

87   & I4/m & $\textrm{\{Z \}} \cong \ZZ_2 \x \ZZ_4$ & $\textrm{\{Y \}} \cong \ZZ_2$ \\

88  & I4$_\textrm{1}$/a &  $\textrm{\{Z \}} \cong \ZZ_2 \x \ZZ_8$ & -\\

97  & I422 & $\textrm{\{Z \}} \cong \ZZ_2 \x \ZZ_4$ & $\textrm{\{Y \}} \cong \ZZ_2 \x \ZZ_2 $\\

98  & I4$_\textrm{1}$22 &  $\textrm{\{Z \}} \cong \ZZ_8$ & $\textrm{\{Y \}} \cong \ZZ_2 \x \ZZ_2$ \\

107  & I4mm & $\textrm{\{Z \}} \cong \ZZ_2 \x \Dn   $ & 
$\textrm{\{Y \}} \cong \ZZ_2 \x \ZZ_2$\\

108  & I4cm & $\textrm{\{Z \}} \cong \ZZ_4 \x \Dn  $ & 
$\textrm{\{Y \}} \cong \ZZ_2 \x \ZZ_2$\\

109  & I4$_\textrm{1}$md & $\textrm{\{Z \}} \cong \ZZ_2^2 \rtimes \ZZ_8 $ & 
$\textrm{\{Y \}} \cong \ZZ_2 \x \ZZ_4 $\\

110  & I4$_\textrm{1}$cd & $\textrm{\{Z \}} \cong \ZZ_2^2 \rtimes \ZZ_8 $  & 
$\textrm{\{Y \}} \cong \ZZ_2 \x \ZZ_4 $\\

119 & I$\overline{\textrm{4}}$m2 & $\textrm{\{Z  \}}\cong \ZZ_2 \x \ZZ_2$ & $\textrm{\{Y \}} \cong \ZZ_2 \x \ZZ_2 $\\

120 & I$\overline{\textrm{4}}$c2 & $\textrm{\{Z \}} \cong \ZZ_2^2 \x \ZZ_4$ & $\textrm{\{Y \}} \cong \ZZ_2 \x \ZZ_2 $\\

121 & I$\overline{\textrm{4}}$2m & $\textrm{\{Z \}} \cong \ZZ_2 \x \ZZ_2$ & $\textrm{\{Y \}} \cong \ZZ_2 \x \ZZ_2 $\\

122 & I$\overline{\textrm{4}}$2d &  $\textrm{\{Z \}} \cong \ZZ_2 \x \ZZ_8$  & $\textrm{\{Y \}} \cong \ZZ_2 \x \ZZ_4$ \\

139 & I4/mmm &  $\textrm{\{Z \}} \cong \ZZ_2 \x \Dn $ & $\textrm{\{Y \}} \cong \ZZ_2 \x \ZZ_2 \x \ZZ_2$ \\

140 & I4/mcm &  $\textrm{\{Z \}} \cong \ZZ_4 \x \Dn $ & $\textrm{\{Y \}} \cong \ZZ_2 \x \ZZ_2 \x \ZZ_2$ \\

141  & I4$_\textrm{1}$amd & $\textrm{\{Z \}} \cong \ZZ_2^2 \rtimes \ZZ_8 $ & 
$\textrm{\{Y \}} \cong \ZZ_2^2 \rtimes \ZZ_4 $ \\

142  & I4$_\textrm{1}$acd & $\textrm{\{Z \}} \cong \ZZ_2^2 \rtimes \ZZ_8 $ & 
$\textrm{\{Y \}} \cong \ZZ_2^2 \rtimes \ZZ_4 $ \\

\bottomrule
\end{tabular*}
\end{minipage}
\end{center}
\end{table*}	 

\begin{table*}[t]
\begin{center}
\begin{minipage}{\textwidth}
\caption{``non-universal" wavevectors' little groups for space groups with Rhombohedral Bravais lattice. Plots of the corresponding Brillouin zones are reported in Fig.\ref{fig:hR}, together with the relations defining the respective cases. } \label{tab:hR}
\begin{tabular*}{\textwidth}{@{\extracolsep{\fill}}rlllcc@{\extracolsep{\fill}}}
\toprule
\multicolumn{2}{@{}c@{}}{Space group} & Case 1 & Case 2 &\\
\midrule
146 	&	R3 &	- & $\textrm{\{P, P1, Q, Z1 \}} \cong \ZZ_3 \x \ZZ_3$\\

148 & R$ \overline{\textrm{3}}$ & - & $\textrm{\{P, P1, Q, Z1 \}} \cong \ZZ_3 \x \ZZ_3$\\

155  & R32 & $\textrm{\{ Q, X, B, B1 \}} \cong \ZZ_2 \x \ZZ_2$ & $\textrm{\{P, P1, Q, Z1 \}} \cong \ZZ_3 \x \ZZ_3$\\

160  & R3m & $\textrm{\{P, P1, P2 \}} \cong \ZZ_2 \x \ZZ_2$ & $\textrm{\{P, P1, Q, Z1  \}} \cong \ZZ_2 \x \textsf{Sym(3)} $\\

161 & R3c & $\textrm{\{P, P1, P2 \}} \cong \ZZ_4$ & $\textrm{\{P, P1, Q, Z1 \}} \cong \textsf{Dic}_3 $  GAP id [12,1]\\

166  & R$\overline{\textrm{3}}$m & $\textrm{\{P, P1, P2, Q, X, B, B1 \}} \cong \ZZ_2 \x \ZZ_2 $ & 
$\textrm{\{P, P1, Q, Z1 \}} \cong \ZZ_2 \x \textsf{Sym(3)} $\\

167 & R$\overline{\textrm{3}}$c & $\textrm{\{P, P1, P2 \}} \cong \ZZ_4 \; \textrm{\{Q, X, B, B1 \}} \cong \ZZ_2 \x \ZZ_2$ & $\textrm{\{P, P1, Q, Z1 \}} \cong \textsf{Dic}_\textsf{3} $ GAP id [12,1]\\

\bottomrule
\end{tabular*}
\end{minipage}
\end{center}
\end{table*}	 

\begin{table*}
\begin{center}
\begin{minipage}{\textwidth}
\caption{``non-universal" wavevectors' little groups for space groups with Cubic face-centred Bravais lattice. Plot of the corresponding Brillouin zone is reported in Fig.\ref{fig:cF}.} \label{tab:cF}
\begin{tabular*}{\textwidth}{@{\extracolsep{\fill}}rlllcc@{\extracolsep{\fill}}}
\toprule
\multicolumn{2}{@{}c@{}}{Space group} & Case 1 \\
\midrule

202 & Fm3 & $\textrm{\{U, K \}} \cong \ZZ_2 $ \\

203  & Fd3 & $\textrm{\{U, K \}} \cong \ZZ_2 \x \ZZ_4 $ \\

209  & F432& $\textrm{\{U, K \}} \cong \ZZ_2 \x \ZZ_2$ \\

210 & F4$_\textrm{1}$32 & $\textrm{\{U, K \}} \cong \ZZ_2 \x \ZZ_2$ \\

216  & F43m & $\textrm{\{U, K \}} \cong \ZZ_2 \x \ZZ_2 $ \\

219 & F43c & $\textrm{\{U \}} \cong \ZZ_2 \x \ZZ_4 \; \textrm{\{K \}} \cong \ZZ_2 \x \ZZ_2$ \\

225 & Fm3m & $\textrm{\{U, K \}}\cong \ZZ_2 \x \ZZ_2 \x \ZZ_2$\\

226 & Fm3c & $\textrm{\{U \}} \cong \ZZ_4 \rtimes \ZZ_4 \; \textrm{\{K \}} \cong \ZZ_2^2 \rtimes \ZZ_4$ \\

227 & Fd3m & $\textrm{\{U \}} \cong \ZZ_2^2 \rtimes \ZZ_4 \; \textrm{\{K \}} \cong \ZZ_4 \x \Dn $\\ 

228 & Fd3c & $\textrm{\{U \}} \cong \ZZ_4 \rtimes \ZZ_4 \; \textrm{\{K \}} \cong \ZZ_4 \x \Dn$  \\ 
\bottomrule
\end{tabular*}
\end{minipage}
\end{center}
\end{table*}


\begin{table*}
\centering
\begin{tabular}{|c|c|c|c|c|}
\midrule
	$(e,000)$ & $(\xb y \zb, \oh \oh \oh)$  & $(e, 010)$  & $(x \yb z, 0 0 \oh)$ & $(\xb \yb \zb, \oh \oh 0)$ \\
\midrule
1 &  1 &  1 &  1 &  1 \\
1 & -1 &  1 & -1 &  1 \\
1 & -1 &  1 &  1 & -1 \\
1 &  1 &  1 & -1 & -1 \\
2 &  0 & -2 &  0 &  0 \\
\midrule
\end{tabular}
\caption{SG15X}
\label{table:SG15X}
\end{table*}

\begin{table}
\centering

\Rotatebox{90}{%
\begin{tabular}{|c|c|c|c|c|c|c|c|c|c|c|c|c|c|}
\midrule
	$(e,000)$ & $(e, \oh \oh 0)$ & $(\xb y z, \tq \oq \tq )$ & $(\xb y z, \oq \tq \tq)$ & $(x \yb z, \oq \tq \tq)$ & $(e, \oh 0 \oh)$ & $(e, \oh 1 \oh)$ & $(e, 0 \oh \oh)$ & $(x \yb z, \oq \oq \oq)$ &$(\xb y z, \tq \tq \oq)$  & $(\xb \yb z, \oh \oh 0)$ & $(\xb y z, \oq \oq \oq)$ & $(\xb \yb z, \oh 0 \oh)$ & $(e, 110)$ \\
\midrule
1  & 1  & 1  & 1  & 1  & 1  & 1  & 1  & 1  & 1  & 1  & 1  & 1  & 1  \\
1  & 1  & -i & -i & -i & -1 & -1 & -1 & i & i & 1  & i & -1 & 1  \\
1  & 1  & -i & -i & i & -1 & -1 & -1 & -i & i & -1 & i & 1  & 1  \\
1  & 1  & 1  & 1  & -1 & 1  & 1  & 1  & -1 & 1  & -1 & 1  & -1 & 1  \\
1  & 1  & i & i & -i & -1 & -1 & -1 & i & -i & -1 & -i & 1  & 1  \\
1  & 1  & i & i & i & -1 & -1 & -1 & -i & -i & 1  & -i & -1 & 1  \\
1  & 1  & -1 & -1 & 1  & 1  & 1  & 1  & 1  & -1 & -1 & -1 & -1 & 1  \\
1  & 1  & -1 & -1 & -1 & 1  & 1  & 1  & -1 & -1 & 1  & -1 & 1  & 1  \\
2  & 0  & -1-i & 1+i & 0  & -2i & 2i & 0  & 0  & -1+i & 0  & 1-i & 0  & -2 \\
2  & 0  & 1-i & -1+i & 0  & 2i & -2i & 0  & 0  & 1+i & 0  & -1-i & 0  & -2 \\
2  & 0  & 1+i & -1-i & 0  & -2i & 2i & 0  & 0  & 1-i & 0  & -1+i & 0  & -2 \\
2  & 0  & -1+i & 1-i & 0  & 2i & -2i & 0  & 0  & -1-i & 0  & 1+i & 0  & -2 \\
2  & -2 & 0  & 0  & 0  & 2  & 2  & -2 & 0  & 0  & 0  & 0  & 0  & 2  \\
2  & -2 & 0  & 0  & 0  & -2 & -2 & 2 & 0  & 0  & 0  & 0  & 0  & 2  \\
\midrule
\end{tabular}
}%
\caption{SG43H}
\label{table:SG43H}
\end{table}

\begin{table}
\centering
\Rotatebox{90}{%
\begin{tabular}{|c|c|c|c|c|c|c|c|c|c|c|c|c|c|}
\midrule
	$(e,000)$ & $(x \yb z, \oq \oq \oq)$ & $(e, 0 \oh \oh)$  & $(x \yb z, \oq \tq \tq)$ & $(e, \oh \oh 0)$ & $(x \yb z, \tq \tq \oq)$  & $(e, 1 \overline{\oh} \oh)$ & $(x \yb z, \tq \oq \tq)$  & $(e, \oh 0 \oh)$ & $(e, 100)$ & $(\xb \yb z, 000)$ & $(\xb y z, \oq \oq \oq)$  & $(\xb y z, \oq \tq \tq)$ & $(\xb \yb z, 0 \oh \oh)$  \\
\midrule
1  & 1  & 1  & 1  & 1  & 1  & 1  & 1  & 1  & 1  & 1  & 1  & 1  & 1  \\
1  & -i & -1 & i & 1  & -i & -1 & i & -1 & 1  & -1 & i & -i & 1  \\
1  & -i & -1 & i & 1  & -i & -1 & i & -1 & 1  & 1  & -i & i & -1 \\
1  & 1  & 1  & 1  & 1  & 1  & 1  & 1  & 1  & 1  & -1 & -1 & -1 & -1 \\
1  & i & -1 & -i & 1  & i & -1 & -i & -1 & 1  & 1  & i & -i & -1 \\
1  & i & -1 & -i & 1  & i & -1 & -i & -1 & 1  & -1 & -i & i & 1  \\
1  & -1 & 1  & -1 & 1  & -1 & 1  & -1 & 1  & 1  & 1  & -1 & -1 & 1  \\
1  & -1 & 1  & -1 & 1  & -1  & 1  & -1  & 1  & 1  & -1 & 1  & 1  & -1  \\
2  & 0  & 2 & 0  & -2 & 0  & 2 & 0  & -2 & 2  & 0  & 0  & 0  & 0  \\
2  & 0  & -2 & 0  & -2 & 0  & -2 & 0  & 2  & 2  & 0  & 0  & 0  & 0  \\
2  & $\sqrt{\textrm{2}}$ & 2  & $\sqrt{\textrm{2}}$ & 0  & $-\sqrt{\textrm{2}}$ & -2 & $-\sqrt{\textrm{2}}$ & 0  & -2 & 0  & 0  & 0  & 0  \\
2  &$\sqrt{\textrm{2}}$i & -2 &  $-\sqrt{\textrm{2}}$i & 0  &  $-\sqrt{\textrm{2}}$i & 2  &$\sqrt{\textrm{2}}$i & 0  & -2 & 0  & 0  & 0  & 0  \\
2  &  $-\sqrt{\textrm{2}}$i & -2 &$\sqrt{\textrm{2}}$i & 0  &$\sqrt{\textrm{2}}$i & 2  &  $-\sqrt{\textrm{2}}$i & 0  & -2 & 0  & 0  & 0  & 0  \\
2  & $-\sqrt{\textrm{2}}$ & 2 & $-\sqrt{\textrm{2}}$ & 0  & $\sqrt{\textrm{2}}$ & -2 & $\sqrt{\textrm{2}}$ & 0  & -2 & 0  & 0  & 0  & 0  \\
\midrule
\end{tabular}
}%
\caption{SG43X}
\label{table:SG43X}
\end{table}

\begin{table*}
\centering
\begin{tabular}{|c|c|c|c|c|c|c|c|c|c|}
\midrule
	$(e,000)$ & $(e, \oh \oh 0)$ & $(e, 1 1 0)$ & $(e, \oh \oh 1)$ & $(x y \zb, \oh \oh \oh)$  & $(x y \zb, 00 \oh)$ & $(x \yb \zb, \oh \oh 0)$ & $(x \yb \zb, 000)$ & $(x \yb z, \oh \oh \oh)$ & $(x \yb z, 0 0 \oh)$ \\
\midrule
1  & 1  & 1  & 1  & 1  & 1  & 1  & 1  & 1  & 1  \\
1  & 1  & 1  & 1  & 1  & 1  & -1 & -1 & -1 & -1 \\
1  & 1  & 1  & 1  & -1 & -1 & 1  & 1  & -1 & -1 \\
1  & 1  & 1  & 1  & -1 & -1 & -1 & -1 & 1 & 1 \\
1  & -1 & 1  & -1 & 1  & -1 & 1 & -1 & -1 & 1 \\
1  & -1 & 1  & -1 & 1  & -1 & -1 & 1  & 1  & -1 \\
1  & -1 & 1  & -1 & -1 & 1 & 1 & -1 & 1 & -1 \\
1  & -1 & 1  & -1 & -1 & 1 & -1 & 1  & -1 & 1 \\
2  & -2i & -2 & 2i & 0  & 0  & 0  & 0  & 0  & 0  \\
2  & 2i & -2 & -2i & 0  & 0  & 0  & 0  & 0  & 0  \\
\midrule
\end{tabular}
\caption{SG63A}
\label{table:SG63A}
\end{table*}

\begin{table*}
\centering
\begin{tabular}{|c|c|c|c|c|c|c|c|c|c|}
\midrule
	$(e,000)$ & $(x \yb \zb, \oh \oh 0)$ & $(x y \zb, \oh \oh \oh)$ & $(x \yb z, 0 0 \oh)$ & $(e, 0 \overline{1} 0)$ & $(x y \zb, \oh \oh \overline{\oh})$ & $(x y \zb, 0 0 \oh)$ & $(x \yb z, \oh \oh \oh)$ & $(x \yb \zb, 000)$ & $(e, \oh \oh 0)$ \\
\midrule
1  & 1  & 1  & 1  & 1  & 1  & 1  & 1  & 1  & 1  \\
1  & -1  & -1  & 1  & 1  & -1  & -1  & 1  & -1  & 1  \\
1  & -1  & -1  & 1  & 1  & -1  & 1  & -1  & 1  & -1  \\
1  & -1  & 1  & -1  & 1  & 1  & -1  & 1  & 1  & -1  \\
1  & -1  & 1  & -1  & 1  & 1  & 1  & -1  & -1  & 1  \\
1  & 1  & -1  & -1  & 1  & -1  & -1  & -1  & 1  & 1  \\
1  & 1  & -1  & -1  & 1  & -1  & 1  & 1  & -1  & -1  \\
1  & 1  & 1  & 1  & 1  & 1  & -1  & -1  & -1  & -1  \\
2  & 0  & -2  & 0  & -2  & 2  & 0  & 0  & 0  & 0  \\
2  & 0  & 2  & 0  & -2  & -2  & 0  & 0  & 0  & 0  \\
\midrule
\end{tabular}
\caption{SG63A1}
\label{table:SG63A1}
\end{table*}

\begin{table*}
\centering
\begin{tabular}{|c|c|c|c|c|c|c|c|c|c|}
\midrule
	$(e,000)$ & $(x \yb z, \oh \oh \oh)$ & $(x y \zb, 0 0 \oh)$ & $(x y \zb, 1 0 \oh)$ & $(x \yb \zb, \oh \oh 0)$ & $(e, 100)$ & $(x y \zb, \oh \oh \oh)$ & $(x \yb \zb, 000)$ & $(x \yb z, 0 0 \oh)$ & $(e, \oh \oh 0)$ \\
\midrule
1  & 1  & 1  & 1  & 1  & 1  & 1  & 1  & 1  & 1  \\
1  & -1  & -1  & -1  & 1  & 1  & -1  & 1  & -1  & 1  \\
1  & -1  & -1  & -1  & 1  & 1  & 1  & -1  & 1  & -1  \\
1  & -1  & 1  & 1  & -1  & 1  & -1  & 1  & 1  & -1  \\
1  & -1  & 1  & 1  & -1  & 1  & 1  & -1  & -1  & 1  \\
1  & 1  & -1  & -1  & -1  & 1  & -1  & -1  & 1  & 1  \\
1  & 1  & -1  & -1  & -1  & 1  & 1  & 1  & -1  & -1  \\
1  & 1  & 1  & 1  & 1  & 1  & -1  & -1  & -1  & -1  \\
2  & 0  & -2  & 2  & 0  & -2  & 0  & 0  & 0  & 0  \\
2  & 0  & 2  & -2  & 0  & -2  & 0  & 0  & 0  & 0  \\
\midrule
\end{tabular}
\caption{SG63X1}
\label{table:SG63X1}
\end{table*}

\begin{table*}
\centering
\begin{tabular}{|c|c|c|c|c|c|c|c|c|c|}
\midrule
	$(e,000)$ & $(e, \oh \oh 0)$ & $(e, \oh \oh 1)$ & $(e, 110)$ & $(x y \zb, 0 \oh \oh)$ & $(x y \zb, \oh 0 \oh)$ & $(x \yb z, \oh 0 \oh)$ & $(x \yb z, 0 \oh \oh)$ & $(x \yb \zb, \oh \oh 0)$ & $(x \yb \zb, 000)$ \\
\midrule
1  & 1  & 1  & 1  & 1  & 1  & 1  & 1  & 1  & 1  \\
1  & 1  & 1  & 1  & 1  & 1  & -1 & -1 & -1 & -1 \\
1  & 1  & 1  & 1  & -1 & -1 & 1  & 1  & -1 & -1 \\
1  & 1  & 1  & 1  & -1 & -1 & -1 & -1 & 1 & 1 \\
1  & -1 & -1 & 1  & 1  & -1 & 1  & -1 & 1  & -1 \\
1  & -1 & -1 & 1  & 1  & -1 & -1 & 1 & -1 & 1 \\
1  & -1 & -1 & 1  & -1 & 1 & 1  & -1 & -1 & 1 \\
1  & -1 & -1 & 1  & -1 & 1 & -1 & 1 & 1 & -1 \\
2  & -2i & 2i & -2 & 0  & 0  & 0  & 0  & 0  & 0  \\
2  & 2i & -2i & -2 & 0  & 0  & 0  & 0  & 0  & 0  \\
\midrule
\end{tabular}
\caption{SG64A}
\label{table:SG64A}
\end{table*}

\begin{table*}
\centering
\begin{tabular}{|c|c|c|c|c|c|c|c|c|c|}
\midrule
$(e,000)$ & $(x y \zb, 0 \oh \oh)$ & $(x y \zb, 1 \oh \oh)$ & $(x y \zb, \oh 0 \oh)$ & $(x \yb \zb, \oh \oh 0)$ & $(x \yb z, \oh 0 \oh)$ & $(x \yb z, 0 \oh \oh)$ & $(x \yb \zb, 000)$ & $(e, \oh \oh 0)$ & $(e, 0 \overline{1} 0)$ \\
\midrule
1  & 1  & 1  & 1  & 1  & 1  & 1  & 1  & 1  & 1  \\
1  & 1  & 1  & 1  & -1 & -1 & -1 & -1 & 1  & 1  \\
1  & 1  & 1  & -1 & 1  & 1  & -1 & -1 & -1 & 1  \\
1  & 1  & 1  & -1 & -1 & -1 & 1 & 1 & -1 & 1  \\
1  & -1 & -1 & 1  & 1  & -1 & 1  & -1 & -1 & 1  \\
1  & -1 & -1 & 1  & -1 & 1 & -1 & 1 & -1 & 1  \\
1  & -1 & -1 & -1 & 1  & -1 & -1 & 1 & 1 & 1  \\
1  & -1 & -1 & -1 & -1 & 1 & 1 & -1 & 1 & 1  \\
2  & -2i & 2i & 0  & 0  & 0  & 0  & 0  & 0  & -2 \\
2  & 2i & -2i & 0  & 0  & 0  & 0  & 0  & 0  & -2 \\
\midrule
\end{tabular}
\caption{SG64A1}
\label{table:SG64A1}
\end{table*}

\begin{table*}
\centering
\begin{tabular}{|c|c|c|c|c|c|c|c|c|c|}
\midrule
	$(e,000)$ & $(x y \zb, \oh 0 \oh)$ & $(x y \zb, \oh 1 \oh)$ & $(x y \zb, 0 \oh \oh)$ & $(e, \oh \oh 0)$ & $(e, 100)$ & $(x \yb \zb, \oh \oh 0)$ & $(x \yb z, 0 \oh \oh)$ & $(x \yb z, \oh 0 \oh)$ & $(x \yb \zb, 000)$ \\
\midrule
1  & 1  & 1  & 1  & 1  & 1  & 1  & 1  & 1  & 1  \\
1  & 1  & 1  & 1  & 1  & 1  & -1 & -1 & -1 & -1 \\
1  & 1  & 1  & -1 & -1 & 1  & 1  & 1  & -1 & -1 \\
1  & 1  & 1  & -1 & -1 & 1  & -1 & -1 & 1 & 1 \\
1  & -1 & -1 & 1  & -1 & 1  & 1 & -1 & 1 & -1 \\
1  & -1 & -1 & 1  & -1 & 1  & -1 & 1  & -1 & 1  \\
1  & -1 & -1 & -1 & 1 & 1  & 1 & -1 & -1 & 1 \\
1  & -1 & -1 & -1 & 1 & 1  & -1 & 1  & 1 & -1 \\
2  & -2i &  2i & 0  & 0  & -2 & 0  & 0  & 0  & 0  \\
2  &  2i & -2i & 0  & 0  & -2 & 0  & 0  & 0  & 0  \\
\midrule
\end{tabular}
\caption{SG64X1}
\label{table:SG64X1}
\end{table*}

\begin{table*}
\centering
\begin{tabular}{|c|c|c|c|c|c|c|c|c|c|}
\midrule
	$(e,000)$ & $(e, \oh \oh 0)$ & $(e, \oh \oh 1)$ & $(x y \zb, \oh \oh \oh)$ & $(x y \zb,0 0 \oh )$ & $(x \yb z, 0 0 \oh)$ & $(x \yb z, \oh \oh \oh)$ & $(x \yb \zb, \oh \oh 0)$ & $(x \yb \zb, 000)$ & $(e,110)$ \\
\midrule
1  & 1  & 1  & 1  & 1  & 1  & 1  & 1  & 1  & 1  \\
1  & 1  & 1  & 1  & 1  & -1 & -1 & -1 & -1 & 1  \\
1  & 1  & 1  & -1 & -1 & 1  & 1  & -1 & -1 & 1  \\
1  & 1  & 1  & -1 & -1 & -1 & -1 & 1 & 1 & 1  \\
1  & -1 & -1 & 1  & -1 & 1  & -1 & 1  & -1 & 1  \\
1  & -1 & -1 & 1  & -1 & -1 & 1 & -1 & 1 & 1  \\
1  & -1 & -1 & -1 & 1 & 1  & -1 & -1 & 1 & 1  \\
1  & -1 & -1 & -1 & 1 & -1 & 1 & 1 & -1 & 1  \\
2  & -2i & 2i & 0  & 0  & 0  & 0  & 0  & 0  & -2 \\
2  & 2i & -2i & 0  & 0  & 0  & 0  & 0  & 0  & -2 \\
\midrule
\end{tabular}
\caption{SG66A}
\label{table:SG66A}
\end{table*}

\begin{table*}
\centering
\begin{tabular}{|c|c|c|c|c|c|c|c|c|c|}
\midrule
	$(e,000)$ & $(x \yb z, 0 0 \oh)$ & $(x \yb \zb, \oh \oh 0)$ & $(x y \zb, \oh \oh \oh)$ & $(e, 100)$ & $(x y \zb, \oh \overline{\oh} \oh)$ & $(x y \zb, 0 0 \oh)$ & $(x \yb \zb, 000)$ & $(x \yb z, \oh \oh \oh)$ & $(e, \oh \oh 0)$ \\
\midrule
1  & 1  & 1  & 1  & 1  & 1  & 1  & 1  & 1  & 1  \\
1  & -1  & -1  & 1  & 1  & 1  & -1  & 1  & 1  & -1  \\
1  & -1  & -1  & 1  & 1  & 1  & 1  & -1  & -1  & 1  \\
1  & -1  & 1  & -1  & 1  & -1  & -1  & 1  & -1  & 1  \\
1  & -1  & 1  & -1  & 1  & -1  & 1  & -1  & 1  & -1  \\
1  & 1  & -1  & -1  & 1  & -1  & -1  & -1  & 1  & 1  \\
1  & 1  & -1  & -1  & 1  & -1  & 1  & 1  & -1  & -1  \\
1  & 1  & 1  & 1  & 1  & 1  & -1  & -1  & -1  & -1  \\
2  & 0  & 0  & -2  & -2  & 2  & 0  & 0  & 0  & 0  \\
2  & 0  & 0  & 2  & -2  & -2  & 0  & 0  & 0  & 0  \\
\midrule
\end{tabular}
\caption{SG66A1}
\label{table:SG66A1}
\end{table*}

\begin{table*}
\centering
\begin{tabular}{|c|c|c|c|c|c|c|c|c|c|}
\midrule
	$(e,000)$ & $(x \yb \zb, \oh \oh 0)$ & $(x y \zb, 0 0 \oh)$ & $(x y \zb, 1 0 \overline{\oh})$ & $(x \yb z, \oh \oh \oh)$ & $(e, 0 \overline{1} 0)$ & $(x y \zb, \oh \oh \oh)$ & $(x \yb \zb, 000)$ &$(e, \oh \oh 0)$  & $(x \yb z, 0 0 \oh)$ \\
\midrule
1  & 1  & 1  & 1  & 1  & 1  & 1  & 1  & 1  & 1  \\
1  & -1  & -1  & -1  & 1  & 1  & -1  & -1  & 1  & 1  \\
1  & -1  & -1  & -1  & 1  & 1  & 1  & 1  & -1  & -1  \\
1  & -1  & 1  & 1  & -1  & 1  & -1  & 1  & -1  & 1  \\
1  & -1  & 1  & 1  & -1  & 1  & 1  & -1  & 1  & -1  \\
1  & 1  & -1  & -1  & -1  & 1  & -1  & 1  & 1  & -1  \\
1  & 1  & -1  & -1  & -1  & 1  & 1  & -1  & -1  & 1  \\
1  & 1  & 1  & 1  & 1  & 1  & -1  & -1  & -1  & -1  \\
2  & 0  & -2  & 2  & 0  & -2  & 0  & 0  & 0  & 0  \\
2  & 0  & 2  & -2  & 0  & -2  & 0  & 0  & 0  & 0  \\
\midrule
\end{tabular}
\caption{SG66X1}
\label{table:SG66X1}
\end{table*}

\begin{table*}
\centering
\begin{tabular}{|c|c|c|c|c|c|c|c|c|c|}
\midrule
	$(e,000)$ & $(x y \zb, \oh 0 0)$ & $(x y \zb, \oh \overline{1} 0)$ & $(x y \zb, 0 \oh 0)$ & $(x \yb \zb, \oh \oh 0)$ & $(x \yb z, 0 \oh 0)$ & $(x \yb z, \oh 0 0)$ & $(x \yb \zb, 000)$ & $(e, \oh \oh 0)$ & $(e, 0 \overline{1} 0)$ \\
\midrule
1  & 1  & 1  & 1  & 1  & 1  & 1  & 1  & 1  & 1  \\
1  & 1  & 1  & 1  & -1 & -1 & -1 & -1 & 1  & 1  \\
1  & 1  & 1  & -1 & 1  & 1  & -1 & -1 & -1 & 1  \\
1  & 1  & 1  & -1 & -1 & -1 & 1 & 1 & -1 & 1  \\
1  & -1 & -1 & 1  & 1  & -1 & 1  & -1 & -1 & 1  \\
1  & -1 & -1 & 1  & -1 & 1 & -1 & 1 & -1 & 1  \\
1  & -1 & -1 & -1 & 1  & -1 & -1 & 1 & 1 & 1  \\
1  & -1 & -1 & -1 & -1 & 1 & 1 & -1 & 1 & 1  \\
2  & -2i & 2i & 0  & 0  & 0  & 0  & 0  & 0  & -2 \\
2  & 2i & -2i & 0  & 0  & 0  & 0  & 0  & 0  & -2 \\
\midrule
\end{tabular}
\caption{SG67A1}
\label{table:SG67A1}
\end{table*}

\begin{table*}
\centering
\begin{tabular}{|c|c|c|c|c|c|c|c|c|c|}
\midrule
	$(e,000)$ & $(x y \zb, \oh 0 0)$ & $(e, 0 \overline{1} 0)$ & $(x y \zb, \oh 1 0)$ & $(x y \zb, 0 \oh 0)$ & $(e, \oh \oh 0)$ & $(x \yb \zb, \oh \oh 0)$ & $(x \yb z, \oh 0 0)$ & $(x \yb \zb, 000)$ & $(x \yb z, 0 \oh 0)$ \\
\midrule
1  & 1  & 1  & 1  & 1  & 1  & 1  & 1  & 1  & 1  \\
1  & 1  & 1  & 1  & 1  & 1  & -1 & -1 & -1 & -1 \\
1  & 1  & 1  & 1  & -1 & -1 & 1  & -1 & -1 & 1  \\
1  & 1  & 1  & 1  & -1 & -1 & -1 & 1  & 1  & -1 \\
1  & -1 & 1  & -1 & 1  & -1 & 1  & 1  & -1 & -1 \\
1  & -1 & 1  & -1 & 1  & -1 & -1 & -1 & 1  & 1  \\
1  & -1 & 1  & -1 & -1 & 1  & 1  & -1 & 1  & -1 \\
1  & -1 & 1  & -1 & -1 & 1  & -1 & 1  & -1 & 1  \\
2  & -2i & -2 & 2i & 0  & 0  & 0  & 0  & 0  & 0  \\
2  & 2i & -2 & -2i & 0  & 0  & 0  & 0  & 0  & 0  \\
\midrule
\end{tabular}
\caption{SG67X1}
\label{table:SG67X1}
\end{table*}

\begin{table*}
\centering
\begin{tabular}{|c|c|c|c|c|c|c|c|c|c|}
\midrule
	$(e,000)$ & $(e, \oh \oh 0)$ & $(e, \oh \overline{\oh} 1)$ & $(x y \zb, \oh 0 \oh)$ & $(x \yb z, 0 \oh \oh)$ & $(x \yb z, \oh 0 \oh)$ & $(x \yb \zb, \oh \oh 0)$ & $(x \yb \zb, 000)$ & $(x y \zb, 0 \oh \oh)$ & $(e, 110)$ \\
\midrule
1  & 1  & 1  & 1  & 1  & 1  & 1  & 1  & 1  & 1  \\
1  & 1  & 1  & 1  & -1 & -1 & -1 & -1 & 1  & 1  \\
1  & 1  & 1  & -1 & 1  & 1  & -1 & -1 & -1 & 1  \\
1  & 1  & 1  & -1 & -1 & -1 & 1 & 1 & -1 & 1  \\
1  & -1 & -1 & 1  & 1  & -1 & 1  & -1 & -1 & 1  \\
1  & -1 & -1 & 1  & -1 & 1 & -1 & 1 & -1 & 1  \\
1  & -1 & -1 & -1 & 1  & -1 & -1 & 1 & 1 & 1  \\
1  & -1 & -1 & -1 & -1 & 1 & 1 & -1 & 1 & 1  \\
2  & -2i & 2i & 0  & 0  & 0  & 0  & 0  & 0  & -2 \\
2  & 2i & -2i & 0  & 0  & 0  & 0  & 0  & 0  & -2 \\
\midrule
\end{tabular}
\caption{SG68A}
\label{table:SG68A_t13}
\end{table*}

\begin{table*}
\centering
\begin{tabular}{|c|c|c|c|c|c|c|c|c|c|}
\midrule
	$(e,000)$ & $(x y \zb, 0 \oh \oh)$ & $(x y \zb, 1 \oh \oh)$ & $(x y \zb \oh 0 \oh)$ & $(x \yb \zb, \oh \oh 0)$ & $(x \yb z, \oh 0 \oh)$ & $(x \yb z, 0 \oh \oh)$ & $(x \yb \zb, 000)$ & $(e, \oh \oh 0)$ & $(e, 0 \overline{1} 0)$ \\
\midrule
1  & 1  & 1  & 1  & 1  & 1  & 1  & 1  & 1  & 1  \\
1  & 1  & 1  & 1  & -1 & -1 & -1 & -1 & 1  & 1  \\
1  & 1  & 1  & -1 & 1  & 1  & -1 & -1 & -1 & 1  \\
1  & 1  & 1  & -1 & -1 & -1 & 1 & 1 & -1 & 1  \\
1  & -1 & -1 & 1  & 1  & -1 & 1  & -1 & -1 & 1  \\
1  & -1 & -1 & 1  & -1 & 1 & -1 & 1 & -1 & 1  \\
1  & -1 & -1 & -1 & 1  & -1 & -1 & 1 & 1 & 1  \\
1  & -1 & -1 & -1 & -1 & 1 & 1 & -1 & 1 & 1  \\
2  & -2i & 2i & 0  & 0  & 0  & 0  & 0  & 0  & -2 \\
2  & 2i & -2i & 0  & 0  & 0  & 0  & 0  & 0  & -2 \\
\midrule
\end{tabular}
\caption{SG68A1}
\label{table:SG68A1}
\end{table*}

\begin{table*}
\centering
\begin{tabular}{|c|c|c|c|c|}
\midrule
	$(e,000)$ & $(x y \zb, 0 \oh \oh)$ & $(e, 0 \overline{11})$ & $(x \yb z, 0 \oh \oh)$ & $(x \yb \zb, 000)$ \\
\midrule
1  & 1  & 1  & 1  & 1  \\
1  & -1  & 1  & -1  & 1  \\
1  & -1  & 1  & 1  & -1  \\
1  & 1  & 1  & -1  & -1  \\
2  & 0  & -2  & 0  & 0  \\
\midrule
\end{tabular}
\caption{SG68X1}
\label{table:SG68X1}
\end{table*}

\begin{table*}
\centering
\begin{tabular}{|c|c|c|c|c|c|c|c|c|c|c|}
\midrule
	$(e,000)$ & $(x \yb \zb, \oh \oq \tq)$ & $(e, 0 \oh \oh)$ & $(x \yb z, \tq \oh \oq)$ & $(x \yb z, \oq 0 \oq)$ & $(e, \oh \oh 0)$ & $(x y \zb, \tq \tq 0)$ & $(x \yb \zb, 0 \oq \oq)$ & $(e, \oh 0 \oh)$ & $(e, 100)$ &$(x y \zb, \oq \oq 0)$  \\
\midrule
1  & 1  & 1  & 1  & 1  & 1  & 1  & 1  & 1  & 1  & 1  \\
1  & 1 & 1  & -1 & 1 & -1 & 1 & -1 & -1 & 1  & -1 \\
1  & 1 & 1  & -1 & -1 & 1  & -1 & 1 & 1  & 1  & -1 \\
1  & 1  & 1  & 1  & -1 & -1 & -1 & -1 & -1 & 1  & 1  \\
1  & -1 & 1  & i & -i & -1 & i & 1 & -1 & 1  & -i \\
1  & -1 & 1  & i & i & 1  & -i & -1 & 1  & 1  & -i \\
1  & -1 & 1  & -i & -i & 1  & i & -1 & 1  & 1  & i \\
1  & -1 & 1  & -i & i & -1 & -i & 1 & -1 & 1  & i \\
2  & 0  & -2 & 0  & 0  & 2 & 0  & 0  & -2 & 2  & 0  \\
2  & 0  & -2 & 0  & 0  & -2 & 0  & 0  & 2  & 2  & 0  \\
4  & 0  & 0  & 0  & 0  & 0  & 0  & 0  & 0  & -4  & 0  \\
\midrule
\end{tabular}
\caption{SG70A}
\label{table:SG70A}
\end{table*}

\begin{table}
\centering
\Rotatebox{90}{%
\begin{tabular}{|c|c|c|c|c|c|c|c|c|c|c|c|c|c|}
\midrule
	$(e,000)$ & $(e, 0 \oh \oh)$  & $(x y \zb, \tq \tq 0)$ & $(x y \zb, \oq \oq 0)$ & $(e, \oh \oh 0)$ & $(x \yb z, \oq 0 \oq)$ & $(x y \zb, \oq \tq \oh)$ & $(x \yb \zb, 0 \tq \tq)$ & $(e, \oh 1 \overline{\oh})$ & $(e, \oh 0 \oh)$ & $(x \yb z, \tq 0 \tq)$ & $(x \yb \zb, \oh \oq \tq)$ & $(e, 0 \overline{1} 0)$ & $(x y \zb, \tq \oq \oh)$ \\
\midrule
1  & 1  & 1  & 1  & 1  & 1  & 1  & 1  & 1  & 1  & 1  & 1  & 1  & 1  \\
1  & 1  & -i & i & -1 & -i & i & -1 & -1 & -1 & i & 1  & 1  & -i \\
1  & 1  & -i & i & -1 & i & i & 1  & -1 & -1 & -i & -1 & 1  & -i \\
1  & 1  & 1  & 1  & 1  & -1 & 1  & -1 & 1  & 1  & -1 & -1 & 1  & 1  \\
1  & 1  & i & -i & -1 & -i & -i & 1  & -1 & -1 & i & -1 & 1  & i \\
1  & 1  & i & -i & -1 & i & -i & -1 & -1 & -1 & -i & 1  & 1  & i \\
1  & 1  & -1 & -1 & 1  & 1  & -1 & -1 & 1  & 1  & 1  & -1 & 1  & -1 \\
1  & 1  & -1 & -1 & 1  & -1 & -1 & 1  & 1  & 1  & -1 & 1  & 1  & -1 \\
2  & 0  & -1-i & -1+i & 0  & 0  & 1-i & 0  & 2i & -2i & 0  & 0  & -2 & 1+i \\
2  & 0  & 1+i & 1-i & 0  & 0  & -1+i & 0  & 2i & -2i & 0  & 0  & -2 & -1-i \\
2  & 0  & -1+i & -1-i & 0  & 0  & 1+i & 0  & -2i & 2i & 0  & 0  & -2 & 1-i \\
2  & 0  & 1-i & 1+i & 0  & 0  & -1-i & 0  & -2i & 2i & 0  & 0  & -2 & -1+i \\
2  & -2 & 0  & 0  & 2 & 0  & 0  & 0  & -2 & -2 & 0  & 0  & 2  & 0  \\
2  & -2 & 0  & 0  & -2 & 0  & 0  & 0  & 2  & 2  & 0  & 0  & 2  & 0  \\
\midrule
\end{tabular}
}%
\caption{SG70C}
\label{table:SG70C}
\end{table}

\begin{table}
\centering
\resizebox{!}{10in}{
\Rotatebox{90}{%
\begin{tabular}{|c|c|c|c|c|c|c|c|c|c|c|c|c|c|c|c|c|c|c|}
\midrule
	$(e,000)$ & $(x y \zb, \tq \tq 0)$  & $(x y \zb, \oq \oq 0)$ & $(e, 1 \overline{\oh} \oh)$  & $(e, \oh \oh 0)$  & $(e, \oh 0 \oh)$ & $(x \yb \zb, \overline{1}, \oq \oq)$ & $(x \yb z, \tq \oh \oq)$  & $(e, 0 \oh \oh)$ & $(x \yb \zb, 0 \oq \oq)$ & $(x \yb \zb, \oh \oq \tq)$ & $(x \yb z, \oq 0 \oq)$ & $(e, \overline{1}00)$ & $(\xb y z, \oh \tq \oq)$ & $(\xb y \zb, \oq 0 \oq)$ & $(\xb y \zb, \oq \oh \tq)$ & $(\xb \yb \zb, 000)$  & $(\xb \yb z, \oq \oq 0)$ & $(\xb \yb z, \oq \tq, \oh)$ \\
\midrule
1  & 1  & 1  & 1  & 1  & 1  & 1  & 1  & 1  & 1  & 1  & 1  & 1  & 1  & 1  & 1  & 1  & 1  & 1  \\
1  & -1  & -1  & 1  & 1  & 1  & -1  & 1  & 1  & -1  & -1  & 1  & 1  & -1  & 1  & 1  & 1  & -1  & -1  \\
1  & -1  & -1  & 1  & 1  & 1  & -1  & 1  & 1  & -1  & -1  & 1  & 1  & 1  & -1  & -1  & -1  & 1  & 1  \\
1  & -1  & -1  & 1  & 1  & 1  & 1  & -1  & 1  & 1  & 1  & -1  & 1  & -1  & 1  & 1  & -1  & 1  & 1  \\
1  & -1  & -1  & 1  & 1  & 1  & 1  & -1  & 1  & 1  & 1  & -1  & 1  & 1  & -1  & -1  & 1  & -1  & -1  \\
1  & 1  & 1  & 1  & 1  & 1  & -1  & -1  & 1  & -1  & -1  & -1  & 1  & -1  & -1  & -1  & 1  & 1  & 1  \\
1  & 1  & 1  & 1  & 1  & 1  & -1  & -1  & 1  & -1  & -1  & -1  & 1  & 1  & 1  & 1  & -1  & -1  & -1  \\
\end{tabular}
}%
}
\caption{SG70X}
\label{table:SG70X}
\end{table}

\begin{table*}
\centering
\begin{tabular}{|c|c|c|c|c|c|c|c|c|c|}
\midrule
	$(e, 000)$ & $(y \xb z, 000)$ & $(\xb \yb z, 000)$ & $(y x z, 000)$  & $(x \yb z, 000)$ & $(e, 110)$ & $(\xb \yb z, 110)$ & $(y \xb z, 110)$ & $(x \yb z, 110)$ & $(y x z, 110)$ \\
\midrule
1 &  1 &  1 &  1 &  1 &  1 &  1 &  1 &  1 &  1 \\
1 & -1 &  1 & -1 &  1 & -1 & -1 &  1 & -1 &  1 \\
1 & -1 &  1 & -1 &  1 &  1 &  1 & -1 &  1 & -1 \\
1 & -1 &  1 &  1 & -1 & -1 & -1 &  1 &  1 & -1 \\
1 & -1 &  1 &  1 & -1 &  1 &  1 & -1 & -1 &  1 \\
1 &  1 &  1 & -1 & -1 & -1 & -1 & -1 &  1 &  1 \\
1 &  1 &  1 & -1 & -1 &  1 &  1 &  1 & -1 & -1 \\
1 &  1 &  1 &  1 &  1 & -1 & -1 & -1 & -1 & -1 \\
2 &  0 & -2 &  0 &  0 & -2 &  2 &  0 &  0 &  0 \\
2 &  0 & -2 &  0 &  0 &  2 & -2 &  0 &  0 &  0 \\
\midrule
\end{tabular}
\caption{SG107Z}
\label{table:SG107Z}
\end{table*}

\begin{table}
\centering
\Rotatebox{90}{%
\resizebox{10in}{!}{
\begin{tabular}{|c|c|c|c|c|c|c|c|c|c|c|c|c|c|c|c|c|c|c|c|}
\midrule
	$(e,000)$ & $(\xb \yb z, \oh \oh \oh)$ & $(e,111)$ & $(\xb \yb z, \overline{\oh} \overline{\oh} \overline{\oh})$ & $(\xb \yb z, 000)$ & $(e, \oh \oh \oh)$ & $(y \xb z, 000)$ & $(y \xb z, \oh \oh \oh)$ & $(\xb \yb z, 0 0 1)$ & $(e,\overline{\oh} \overline{\oh} \overline{\oh})$ & $(y \xb z, 1 \overline{1} 1)$ & $(y \xb z, \tm \overline{\oh} \tm)$ & $(y x z, \oh \oh 0)$ & $(y x z, 0 0 \oh)$ & $(y x z, \oh \oh 1)$ & $(y x z, 1 1 \tm)$ & $(x \yb z, \oh \oh 1)$ & $(x \yb z, \oh \oh 0)$ & $(x \yb z, 0 0 \oh)$ & $(x \yb z, 1 \overline{1} \tm)$ \\
\midrule
1 &  1 &  1 &  1 & 1 &  1 &  1 &  1 &  1 &  1 &  1 &  1 &  1 &  1 &  1 	&  1 &  1 &  1 &  1 &  1 \\
1 & -i & -1 &  i & 1 & -i &  1 & -i & -1 &  i & -1 &  i &  1 & -i & -1 	&  i & -1 &  1 & -i &  i \\
1 & -i & -1 &  i & 1 & -i &  1 & -i & -1 &  i & -1 &  i & -1 &  i &  1 	& -i &  1 & -1 &  i & -i \\
1 & -i & -1 &  i & 1 & -i & -1 &  i & -1 &  i &  1 & -i &  1 & -i & -1 	&  i &  1 & -1 &  i & -i \\
1 & -i & -1 &  i & 1 & -i & -1 &  i & -1 &  i &  1 & -i & -1 &  i &  1 	& -i & -1 &  1 & -i &  i \\
1 &  1 &  1 &  1 & 1 &  1 &  1 &  1 &  1 &  1 &  1 &  1 & -1 & -1 & -1 	& -1 & -1 & -1 & -1 & -1 \\
1 &  1 &  1 &  1 & 1 &  1 & -1 & -1 &  1 &  1 & -1 & -1 &  1 &  1 &  1 	&  1 & -1 & -1 & -1 & -1 \\
\end{tabular}
}
}
\caption{SG108Z}
\label{table:SG108Z}
\end{table}

\begin{table}
\centering
\Rotatebox{90}{%
\resizebox{10in}{!}{
\begin{tabular}{|c|c|c|c|c|c|c|c|c|c|c|c|c|c|c|c|c|c|c|c|}
\midrule
	$(e,000)$ & $(y \xb z, \oh 0 \tq)$ & $(y \xb z, 0 \oh \oq)$ & $(y \xb z, \oh 0 \frac{7}{4})$ & $(y \xb z, 1 \overline{\oh} \frac{5}{4})$ & $(x \yb z, 000)$ & $(y x z, \oh 0 \tq)$ & $(x \yb z, \oh \overline{\oh} \tm)$ & $(y x z, 0 \oh \oq)$ & $(y x z, 1 \overline{\oh} \frac{5}{4})$ & $(x \yb z, 101)$ & $(y x z, \oh 0 \frac{7}{4})$ & $(x \yb z, \oh \oh \oh)$ & $(e, \oh \overline{\oh} \tm)$ & $(e,\oh \oh \oh)$ & $(\xb \yb z, \oh \overline{\oh} \tm)$ & $(\xb \yb z, \oh \oh \oh)$ & $(\xb \yb z, 000)$ & $(\xb \yb z, 101)$ & $(e, 001)$ \\
\midrule
1 & 1  & 1  & 1  & 1  & 1  & 1  & 1  & 1  & 1  & 1  & 1  & 1  & 1  & 1  & 1  & 1  & 1  & 1  & 1  \\
1 &  ${\zeta_8}^5$ & ${\zeta_8}^7$ & $\zeta_8$ &  ${\zeta_8}^3$ & 1  &  ${\zeta_8}^5$ & i & ${\zeta_8}^7$ &  ${\zeta_8}^3$ & -1 & $\zeta_8$ & -i & i & -i & i & -i & 1  & -1 & -1 \\
1 &  ${\zeta_8}^5$ & ${\zeta_8}^7$ & $\zeta_8$ &  ${\zeta_8}^3$ & -1 & $\zeta_8$ & -i &  ${\zeta_8}^3$ & ${\zeta_8}^7$ & 1  &  ${\zeta_8}^5$ & i & i & -i & i & -i & 1  & -1 & -1 \\
1 & -i & i & -i & i & 1  & -i & -1 & i & i & 1  & -i & -1 & -1 & -1 & -1 & -1 & 1  & 1  & 1  \\
1 & -i & i & -i & i & -1 & i & 1  & -i & -i & -1 & i & 1  & -1 & -1 & -1 & -1 & 1  & 1  & 1  \\
1 & ${\zeta_8}^7$ &  ${\zeta_8}^5$ &  ${\zeta_8}^3$ & $\zeta_8$ & 1  & ${\zeta_8}^7$ & -i &  ${\zeta_8}^5$ & $\zeta_8$ & -1 &  ${\zeta_8}^3$ & i & -i & i & -i & i & 1  & -1 & -1 \\
1 & ${\zeta_8}^7$ &  ${\zeta_8}^5$ &  ${\zeta_8}^3$ & $\zeta_8$ & -1 &  ${\zeta_8}^3$ & i & $\zeta_8$ &  ${\zeta_8}^5$ & 1  & ${\zeta_8}^7$ & -i & -i & i & -i & i & 1  & -1 & -1 \\
\end{tabular}
}}
\caption{SG109Z}
\label{table:SG109Z}
\end{table}

\begin{table*}
\centering
\Rotatebox{90}{%
\resizebox{10in}{!}{
\begin{tabular}{|c|c|c|c|c|c|c|c|c|c|c|c|c|c|c|c|c|c|c|c|}
\midrule
	$(e,000)$ & $(\xb \yb z, \oh \oh \oh)$ & $(e, 001)$ & $(\xb \yb z, \oh \oh \tm)$ & $(\xb \yb z, 000)$ & $(e, \oh \oh \oh)$ & $(y \xb z, 101)$ & $(e, \oh \oh \tm)$ & $(y \xb z, 000)$ & $(y \xb z, \oh \oh \oh)$ & $(\xb \yb z, 101)$ & $(y \xb z, \oh \oh \tm)$ & $(x \yb z, \oh \oh 1)$ & $(x \yb z, 0 0 \tm)$ & $(x \yb z, \oh \oh 0)$ & $(x \yb z, 0 0 \oh)$ & $(y x z, \oh \oh 1)$ & $(y x z, \oh \oh 0)$ & $(y x z, 0 0 \oh)$ & $(y x z, 0 0 \tm)$ \\
\midrule
1  & 1  & 1  & 1  & 1  & 1  & 1  & 1  & 1  & 1  & 1  & 1  & 1  & 1  & 1  & 1  & 1  & 1  & 1  & 1  \\
1  & -i & -1 & i & 1  & -i & 1  & i & -1 & i & -1 & -i & 1  & -i & -1 & i & -1 & 1  & -i & i \\
1  & -i & -1 & i & 1  & -i & 1  & i & -1 & i & -1 & -i & -1 & i & 1 & -i & 1 & -1 & i & -i \\
1  & -i & -1 & i & 1  & -i & -1 & i & 1 & -i & -1 & i & 1  & -i & -1 & i & 1 & -1 & i & -i \\
1  & -i & -1 & i & 1  & -i & -1 & i & 1 & -i & -1 & i & -1 & i & 1 & -i & -1 & 1 & -i & i \\
1  & 1  & 1  & 1  & 1  & 1  & 1  & 1  & 1  & 1  & 1  & 1  & -1 & -1 & -1 & -1 & -1 & -1 & -1 & -1 \\
1  & 1  & 1  & 1  & 1  & 1  & -1 & 1  & -1 & -1 & 1  & -1 & 1  & 1  & 1  & 1  & -1 & -1 & -1 & -1 \\
\end{tabular}
}}
\caption{SG140Z}
\label{table:SG140Z}
\end{table*}

\begin{table*}
\centering
\begin{tabular}{|c|c|c|c|c|c|c|c|c|c|}
\midrule
	$(e,000)$ & $(e, \oh \overline{\oh} \tm)$ & $(e, \oh \oh \oh)$ & $(\yb \xb z, 0 \oh \oq)$ & $(e,100)$ & $(\yb \xb z, \oh 0 \tq)$ & $(\yb \xb z, 000)$ & $(x y \zb, 0 \oh \oq)$ & $(\yb \xb \zb, 0 \overline{1} 0)$ & $(x y \zb, \oh 0 \tq)$ \\
\midrule
1  & 1  & 1  & 1  & 1  & 1  & 1  & 1  & 1  & 1  \\
1  & 1 & 1  & -1 & 1 & -1 & 1  & -1 & 1 & -1 \\
1  & 1 & 1  & -1 & 1 & -1 & -1 & 1 & -1 & 1 \\
1  & 1  & 1  & 1  & 1  & 1  & -1 & -1 & -1 & -1 \\
1  & -1 & 1  &-i & -1 & i & 1  &-i & -1 & i \\
1  & -1 & 1  &-i & -1 & i & -1 & i & 1 &-i \\
1  & -1 & 1  & i & -1 &-i & 1  & i & -1 &-i \\
1  & -1 & 1  & i & -1 &-i & -1 &-i & 1 & i \\
2  & 2  & -2 & 0  & -2 & 0  & 0  & 0  & 0  & 0  \\
2  & -2 & -2 & 0  & 2  & 0  & 0  & 0  & 0  & 0  \\
\midrule
\end{tabular}
\caption{SG141Y}
\label{table:SG141Y}
\end{table*}

\begin{table*}
\centering
\begin{tabular}{|c|c|c|c|c|c|}
\midrule
	$(e,000)$ & $(e,110)$ & $(y z x, 110)$ & $(y z x, 000)$ & $(z y x, 000)$ & $(z y x, 110)$ \\
\midrule
1  & 1  & 1  & 1  & 1  & 1  \\
1  & -1  & -1  & 1  & -1  & 1  \\
1  & -1  & -1  & 1  & 1  & -1  \\
1  & 1  & 1  & 1  & -1  & -1  \\
2  & -2  & 1  & -1  & 0  & 0  \\
2  & 2  & -1  & -1  & 0  & 0  \\
\midrule
\end{tabular}
\caption{SG160P}
\label{table:SG160P}
\end{table*}

\begin{table*}
\centering
\begin{tabular}{|c|c|c|c|c|c|}
\midrule
 $(e,000)$ & $(e,111)$  & $(y z x, 111)$ & $(z y x, \tm \tm \tm)$ & $(z y x, \oh \oh \oh)$ & $(y z x, 000)$ \\
\midrule
1  & 1  & 1  & 1  & 1  & 1  \\
1  & 1 & 1 & -1 & -1 & 1  \\
1  & -1 & -1 & i & -i & 1  \\
1  & -1 & -1 & -i & i & 1  \\
2  & 2 & -1 & 0  & 0  & -1 \\
2  & -2 & 1  & 0  & 0  & -1 \\
\midrule
\end{tabular}
\caption{SG161P}
\label{table:SG161P}
\end{table*}

\begin{table*}
\centering
\begin{tabular}{|c|c|c|c|c|c|c|c|c|c|}
\midrule
	$(e,000)$ & $(y x \zb, \oh 1 0)$ & $(y x \zb, \oh 0 0)$ & $(e, \oh \oh 0)$ & $(e, \tm \oh 0)$ & $(e, 100)$ & $(x y \zb, \oh 1 \oh)$ & $(y x z, \oh \oh \oh)$ & $(y x z, 0 0 \oh)$ & $(x y \zb, 0 \oh \oh)$ \\
\midrule
1 & 1 & 1 & 1 & 1 & 1 & 1 & 1 & 1 & 1 \\
1 & i & i & -1 & 1 & -1 & i & -1 & 1 & i \\
1 & i & i & -1 & 1 & -1 & i & 1 & -1 & i \\
1 & 1 & 1 & 1 & 1 & 1 & -1 & -1 & -1 & -1 \\
1 & i & i & -1 & 1 & -1 & i & 1 & -1 & i \\
1 & i & i & -1 & 1 & -1 & i & -1 & 1 & i \\
1 & -1 & -1 & 1 & 1 & 1 & 1 & -1 & -1 & 1 \\
1 & -1 & -1 & 1 & 1 & 1 & -1 & 1 & 1 & -1 \\
2 & 0 & 0 & 2 & -2 & -2 & 0 & 0 & 0 & 0 \\
2 & 0 & 0 & -2 & -2 & 2 & 0 & 0 & 0 & 0 \\
\midrule
\end{tabular}
\caption{SG226K}
\label{table:SG226K}
\end{table*}

\begin{table*}
\centering
\begin{tabular}{|c|c|c|c|c|c|c|c|c|c|}
\midrule
	$(e,000)$ & $(z y x, 0 \oh 0)$ & $(z y x, 1 \oh 0)$ & $(e, \oh 1 \oh)$ & $(z \yb x, 0 0 \oh)$ & $(x \yb z, 0 \oh \oh)$ & $(z \yb x, \oh 0 0)$ & $(e, \oh 0 \oh)$ & $(x \yb z, \oh \oh 0)$ & $(e,001)$ \\
\midrule
1 & 1 & 1 & 1 & 1 & 1 & 1 & 1 & 1 & 1 \\
1 & -i & i & -1 & 1 & i & -1 & 1 & -i & -1 \\
1 & -i & i & -1 & -1 & -i & 1 & 1 & i & -1 \\
1 & 1 & 1 & 1 & -1 & -1 & -1 & 1 & -1 & 1 \\
1 & i & -i & -1 & 1 & -i & -1 & 1 & i & -1 \\
1 & i & -i & -1 & -1 & i & 1 & 1 & -i & -1 \\
1 & -1 & -1 & 1 & 1 & -1 & 1 & 1 & -1 & 1 \\
1 & -1 & -1 & 1 & -1 & 1 & -1 & 1 & 1 & 1 \\
2 & 0 & 0 & 2 & 0 & 0 & 0 & -2 & 0 & -2 \\
2 & 0 & 0 & -2 & 0 & 0 & 0 & -2 & 0 & 2 \\
\midrule
\end{tabular}
\caption{SG226U}
\label{table:SG226U}
\end{table*}

\begin{table}
\centering
\Rotatebox{90}{%
\resizebox{8in}{!}{
\begin{tabular}{|c|c|c|c|c|c|c|c|c|c|c|c|c|c|}
\midrule
	$(e,000)$ & $(z \yb x, \oq \oq \oq)$ & $(z y x, \oh 0 \oh)$ & $(z y x, \oh \oh 0)$ & $(e, 0 \oh \oh)$ & $(e, \oh 0 \oh)$ & $(e, 010)$ & $(z \yb x, \tq \frac{5}{4} \tq)$ & $(x \yb z, \oq \oq \oq)$ & $(x \yb z, \tq \oq \tq)$ & $(x \yb z, \oq \tq \tq)$ & $(e, \tm 0 \oh)$ & $(z y x, 000)$ & $(z \yb x, \oq \tq \tq)$ \\
\midrule
1 & 1 & 1 & 1 & 1 & 1 & 1 & 1 & 1 & 1 & 1 & 1 & 1 & 1 \\
1 & 1 & 1 & -1 & -1 & 1 & 1 & 1 & 1 & 1 & -1 & 1 & 1 & -1 \\
1 & 1 & -1 & 1 & -1 & 1 & 1 & 1 & -1 & -1 & 1 & 1 & -1 & -1 \\
1 & 1 & -1 & -1 & 1 & 1 & 1 & 1 & -1 & -1 & -1 & 1 & -1 & 1 \\
1 & -1 & 1 & 1 & 1 & 1 & 1 & -1 & -1 & -1 & -1 & 1 & 1 & -1 \\
1 & -1 & 1 & -1 & -1 & 1 & 1 & -1 & -1 & -1 & 1 & 1 & 1 & 1 \\
1 & -1 & -1 & 1 & -1 & 1 & 1 & -1 & 1 & 1 & -1 & 1 & -1 & 1 \\
1 & -1 & -1 & -1 & 1 & 1 & 1 & -1 & 1 & 1 & 1 & 1 & -1 & -1 \\
2 & 0 & 0 & 0 & 0 & -2 & 2 & 0 & -2i & 2i & 0 & -2 & 0 & 0 \\
2 & 0 & 0 & 0 & 0 & -2 & 2 & 0 & 2i & -2i & 0 & -2 & 0 & 0 \\
2 & 0 & 2 & 0 & 0 & -2 & -2 & 0 & 0 & 0 & 0 & 2 & -2 & 0 \\
2 & 0 & -2 & 0 & 0 & -2 & -2 & 0 & 0 & 0 & 0 & 2 & 2 & 0 \\
2 & 2 & 0 & 0 & 0 & 2 & -2 & -2 & 0 & 0 & 0 & -2 & 0 & 0 \\
2 & -2 & 0 & 0 & 0 & 2 & -2 & 2 & 0 & 0 & 0 & -2 & 0 & 0 \\
\midrule
\end{tabular}
}}
\caption{SG227U}
\label{table:SG227U}
\end{table}

\begin{table}
\centering
\Rotatebox{90}{%
\resizebox{10in}{!}{
\begin{tabular}{|c|c|c|c|c|c|c|c|c|c|c|c|c|c|c|c|c|c|c|c|}
\midrule
	$(e,000)$ & $(x y \zb, \tq \tq \tq)$ & $(x y \zb, \oq \oq \tq)$ & $(x y \zb, \frac{7}{4} \tq \overline{\oq})$ & $(x y \zb, \frac{5}{4} \oq \tq)$ & $(x y \zb, \oq \tq \oq)$ & $(x y \zb, \tq \frac{5}{4} \oq)$ & $(e, 0 \oh \oh)$ & $(e, 1 \oh \oh)$ & $(y x z, 1 \oh 0)$ & $(y x \zb, \tq \tq \oq)$ & $(y x z, \oh \oh \oh)$ & $(y x \zb, \frac{5}{4} \tq \tq)$ & $(y x \zb, \tq \oq \tq)$ & $(y x z, \oh 0 0)$ & $(y x \zb, \oq \oq \oq)$ & $(y x z, 0 0 \oh)$ & $(e, 101)$ & $(e, \tm \oh 0)$ & $(e, \oh \oh 0)$ \\
\midrule
1 & 1 & 1 & 1 & 1 & 1 & 1 & 1 & 1 & 1 & 1 & 1 & 1 & 1 & 1 & 1 & 1 & 1 & 1 & 1 \\
1 & -i & i & i & -i & 1 & -1 & -i & i & -i & i & 1 & 1 & -1 & i & -i & -1 & -1 & 1 & -1 \\
1 & -i & i & i & -i & 1 & -1 & -i & i & i & -i & -1 & -1 & 1 & -i & i & 1 & -1 & 1 & -1 \\
1 & -i & i & i & -i & -1 & 1 & i & -i & -i & -i & -1 & 1 & -1 & i & i & 1 & -1 & 1 & -1 \\
1 & -i & i & i & -i & -1 & 1 & i & -i & i & i & 1 & -1 & 1 & -i & -i & -1 & -1 & 1 & -1 \\
1 & 1 & 1 & 1 & 1 & 1 & 1 & 1 & 1 & -1 & -1 & -1 & -1 & -1 & -1 & -1 & -1 & 1 & 1 & 1 \\
1 & 1 & 1 & 1 & 1 & -1 & -1 & -1 & -1 & 1 & -1 & -1 & 1 & 1 & 1 & -1 & -1 & 1 & 1 & 1 \\
\end{tabular}
}}
\caption{SG228K}
\label{table:SG228K}
\end{table}

\clearpage

\begin{table}[p]
\begin{center}
\begin{minipage}{\textwidth}
\caption{Generators for Abelian local symmetry groups of space groups with Monoclinic Bravais lattices. Plots of the corresponding Brillouin zones are reported in Fig.\ref{fig:mP} (primitive) and Figs.\ref{fig:mC1-3}-\ref{fig:mC4-5} (base-centred), together with the relations defining the respective cases.} \label{tab:Gen_m}
\begin{tabular*}{\textwidth}{@{\extracolsep{\fill}}rlllcc@{\extracolsep{\fill}}}
\toprule
\multicolumn{2}{@{}c@{}}{Space group} & Cases 1, 2 & Cases 3, 4 & Case 5 \\
\midrule
5 	& 	C2	 &$\textrm{X,I}: \; (\xb y \zb, \zz)$& $\textrm{F}: \; (\xb y \zb, \zz)$& $\textrm{I}: \; (\xb y \zb,\zz) $\\

6 	&	Pm	& $\textrm{M, H, M1, H1}: \; (x \yb z, \zz) $ & & \\ 

7 & Pc   &$\textrm{M, H, M1, H1}: \; (x \yb z, 0 0 \oh) $  & & \\

8 & Cm	 &$\textrm{F, F1,X}: \; (x \yb z, \zz)$ & $\textrm{Y, F, H}: \; (x \yb z, \zz)$& $\textrm{Y, F, H}: \; (x \yb z,\zz)$\\

9 & Cc  & $\textrm{F, X}: \; (x \yb z, \zz) \;\textrm{F1}: \; (x \yb z, 0 0 \oh)$ & $\textrm{Y}: \; (x \yb z, \zz)\; \textrm{F,H}: \; (x \yb z, 0 0 \oh) $ & $\textrm{Y, F, H}: \; (x \yb z, \zz) $ \\

10 & P2/m	& $\textrm{M, H, M1, H1}: \; (x \yb z, \zz) $ & & \\

11 & P2$_{\textrm{1}}$/m	& $\textrm{M, H, M1, H1}: \; (x \yb z, \zz) $ & & \\

{\multirow{2}{*}{12}} & {\multirow{2}{*}{C2/m}} & $\textrm{F, F1}: \; (x \yb z, \zz) \; \textrm{I}: \; (\xb y \zb, \zz) $ & $\textrm{F}: \; (x \yb z,\zz), (\xb y \zb, \zz)$ & $\textrm{Y, F, H}: \; (x \yb z, \zz) \;$\\ & &$ \textrm{X}: \; (x \yb z,\zz), (\xb y \zb,\zz)$ & $\textrm{Y, H}: \; (x \yb z, \zz) \; $ & $ \textrm{I}: \; (\xb y \zb, \zz)$ \\

13& P2/c &$\textrm{M, H, M1, H1}: \; (x \yb z, 0 0 \oh) $  & & \\

14& P2$_\textrm{1}$/c &$\textrm{M, H, M1, H1}: \; (x \yb z, 0 \oh \oh) $  & & \\

{\multirow{2}{*}{15}} & {\multirow{2}{*}{C2/c}} & $\textrm{F}: \; (x \yb z, \zz)  \; \textrm{I}: \; (\xb y \zb, \zz) $ & $\textrm{F}: \; (x \yb z, \oh \oh \oh), (\xb y \zb, \oh \oh \oh),(e,111)   $ & $ \textrm{F}: \; (x \yb z,\zz) \;\textrm{I}: (\xb y \zb,\zz)$ \\
 & & $ \textrm{F1}: \; (x \yb z, 0 0 \oh) $  &  $\textrm{Y}: \; (x \yb z, 0 0 \oh) \; \textrm{H}: \; (x \yb z,\zz)$ & $\textrm{Y, H}: \; (x \yb z, 0 0 \oh) $ \\

\bottomrule
\end{tabular*}
\end{minipage}
\end{center}
\end{table}

\begin{table}[p]
\begin{center}
\begin{minipage}{\textwidth}
\caption{Generators for Abelian local symmetry groups of space groups with Orthorhombic base-centred and body-centred Bravais lattices. Plots of the corresponding Brillouin zones are reported in Fig.\ref{fig:oS}. } \label{tab:Gen_oS}
\begin{tabular*}{\textwidth}{@{\extracolsep{\fill}}rlllcc@{\extracolsep{\fill}}}
\toprule
\multicolumn{2}{@{}c@{}}{Space group} & Case 1 \\
\midrule
20 	&	C222$_\textrm{1}$	&$\textrm{A, A1, X, X1}: \; (x \yb \zb, \oh \oh 0) $ \\

21	& C222	 & $\textrm{A, A1, X, X1}: \; (x \yb \zb, \zz)$\\

23	& I222	 & $\textrm{X} \; (x \yb \zb, \zz) \; \textrm{Y}: \; (\xb y \zb, \zz)$\\

24	& I2$_\textrm{1}$2$_\textrm{1}$2$_\textrm{1}$	 & $\textrm{X}: \; (x \yb \zb,\oh \oh 0) \; \textrm{Y}: \; (\xb y \zb, 0 \oh \oh)$\\

35 	& 	Cmm2 &	$\textrm{A,X, A1, X1}: \; (x \yb z, \zz)$\\

36	&	Cmc2$_\textrm{1}$ & $\textrm{A, A1}: \; (x \yb z, 0 0 \oh) \; \textrm{X, X1}: \; (x \yb z, \oh \oh \oh)$\\

37  & 	Ccc2 & $\textrm{A, A1}: \; (x \yb z, 0 0 \oh) \; \textrm{X, X1}: \; (x \yb z, \oh \oh \oh) $\\
\bottomrule
\end{tabular*}
\end{minipage}
\end{center}
\end{table}

\begin{table}[p]
\begin{center}
\begin{minipage}{\textwidth}
\caption{Generators for Abelian local symmetry groups of space groups with Orthorhombic face-centred Bravais lattice. Plots of the corresponding Brillouin zones are reported in Fig.\ref{fig:oF}, together with the relations defining the respective cases.} \label{tab:Gen_oF}
\begin{tabular*}{\textwidth}{@{\extracolsep{\fill}}rlllcc@{\extracolsep{\fill}}}
\toprule
\multicolumn{2}{@{}c@{}}{Space group} & Cases 1,3 & Case 2 &\\
\midrule

22 	& F222 & $\textrm{A}: \; (x \yb \zb, \zz) \; \textrm{X}: (x \yb \zb, \zz),(\xb y \zb, \zz)$ & $\textrm{C}: \; (x \yb \zb,\zz)\; \textrm{D}: \; (\xb y \zb,\zz) \; \textrm{H}: (\xb \yb z, \zz)$ \\

42	& Fmm2 & $\textrm{A}: \; (x \yb z, \zz) \; \textrm{X}: (x \yb z, \zz),(\xb y z, \zz)$ & $\textrm{C}: \; (x \yb z, \zz) \; \textrm{D}: \; (\xb y z, \zz) \; \textrm{H}: \; (\xb y z, \zz),(x \yb z, \zz)$\\

43  & Fdd2 & $\textrm{A}: \;(x \yb z, \tq \tq \oq), (e,\oh 0 \oh)  $ & $\textrm{C}: \; (x \yb z, \tq \oq \tq) \; \textrm{D}: \; (\xb y z, \oq \tq \tq)$\\

{\multirow{2}{*}{69}}	& {\multirow{2}{*}{Fmmm}}	& $\textrm{A}: \; (x \yb z,\zz), (x y \zb,\zz) \; $ & $\textrm{C}: \; (x \yb z,\zz), (x y \zb, \zz)
\textrm{D}: \; (\xb y z,\zz), (x y \zb, \zz)
$\\ 
&& $\textrm{X}: \; (\xb \yb \zb,\zz),(\xb \yb z,\zz), (x \yb \zb, \zz)$ &  $\textrm{H}: \; (\xb y z,\zz), (x \yb z, \zz)$\\

\bottomrule
\end{tabular*}
\end{minipage}
\end{center}
\end{table}	 

\begin{table}[p]
\begin{center}
\begin{minipage}{\textwidth}
\caption{Generators for Abelian local symmetry groups of space groups with Tetragonal body-centred Bravais lattice. Plots of the corresponding Brillouin zones are reported in Fig.\ref{fig:tI}, together with the relations defining the respective cases.} \label{tab:Gen_tI}
\begin{tabular*}{\textwidth}{@{\extracolsep{\fill}}rlllcc@{\extracolsep{\fill}}}
\toprule
\multicolumn{2}{@{}c@{}}{Space group} & Case 1 & Case 2 &\\
\midrule
79  	&	I4 &	$\textrm{Z}: \; (\xb \yb z, \zz), (\yb x z, 1 1 0)$ & -\\

80		&	I4$_\textrm{1}$ &	$\textrm{Z}: \; (y \xb z, \oh 0 \tq), (\xb \yb z,\zz) \; \textrm{Z1}: \; (\yb x z, \oh 0 \tq)$& - \\

82		&	I$\overline{\textrm{4}}$ &
$\textrm{Z}: \; (\xb \yb z, \zz)$	&	-\\

87   & I4/m & $\textrm{Z}: \;  (y \xb z, \zz),  (e, 1 1 0) \; \textrm{Z1}: \; (\yb x z, \zz),(\xb \yb z, 110)$ &  Y:$\; (x y \zb, \zz)$\\

88	& I4$_\textrm{1}$/a & Z:$\; (y \xb z, \frac{5}{4} \oq \frac{5}{4}), (\xb \yb z, 1 \oh 1)$
&- \\	

97  & I422 & $\textrm{Z}: \;   (y \xb z, \zz), \; (e, 1 1 0)$ & $\textrm{Y}: \; (\yb \xb \zb,\zz), \; (e,110) $\\

98  & I4$_\textrm{1}$22 & Z:$\; (y \xb z,0 \oh \oq)$ & $\textrm{Y}: \; (\yb \xb \zb,\oh \oh \oh), \; (e,110)$ \\

107  & I4mm & - & 
$\textrm{Y}: \; (\yb \xb z,\zz), \; (e,110)$\\

108  & I4cm & - & 
$\textrm{Y}: \; (\yb \xb z,\zz), \; (e,110)$\\

109  & I4$_\textrm{1}$md & - & 
$\textrm{Y}: \; (\yb \xb z, \oh 0 \tq),(e,0 \overline{1} 1 ) $\\

110  & I4$_\textrm{1}$cd & - & 
$\textrm{Y}: \; (\yb \xb z, 0 \oh \tq), (e,\overline{1} 0 1) $\\

119 & I$\overline{\textrm{4}}$m2 & Z:$(\xb y z,\zz),(x \yb z, \zz)$  & $\textrm{Y}: \;  (\yb \xb \zb, \zz), \; (e, 110) $\\

120 & I$\overline{\textrm{4}}$c2 & Z:$(\xb \yb z, \oh \oh \oh),(\xb \yb z,\zz),(x \yb z, \oh \overline{\oh} 1)$ & $\textrm{Y}: \;  (\yb \xb \zb, \oh \oh 0), \; (e, 110) $\\

121 & I$\overline{\textrm{4}}$2m &   & $\textrm{Y}: \;  (\yb \xb z, \zz), \; (e, 110) $\\

122 & I$\overline{\textrm{4}}$2d &  Z:$(\yb \xb z, 0 \oh \oq), (\xb \yb z,001)$ & $\textrm{Y}: \; (\yb \xb z, \oh 0 \tq), (e,0 \overline{1} 1)$ \\

139 & I4/mmm &  - & $\textrm{Y}:  \; (\yb \xb \zb, \zz), (e,110), (x y \zb, 110)$ \\

140	&	I4/mcm	& - & Y:$\; (\yb \xb \zb, \oh \oh 0),(e, 110),(x y \zb, 110)$\\

\bottomrule
\end{tabular*}
\end{minipage}
\end{center}
\end{table}

\begin{table}[p]
\begin{center}
\begin{minipage}{\textwidth}
\caption{Generators for Abelian local symmetry groups of space groups with Rhombohedral Bravais lattice. Plots of the corresponding Brillouin zones are reported in Fig.\ref{fig:hR}, together with the relations defining the respective cases.} \label{tab:Gen_hR}
\begin{tabular*}{\textwidth}{@{\extracolsep{\fill}}rlllcc@{\extracolsep{\fill}}}
\toprule
\multicolumn{2}{@{}c@{}}{Space group} & Case 1 & Case 2 &\\
\midrule
146 	&	R3 &	- & $\textrm{P, P1, Q, Z1}: \; (y z x,\zz), (e, 120) $\\

148 & R$ \overline{\textrm{3}}$ & - & $\textrm{P, P1, Q, Z1}: \; (y z x, \zz), \; (e, 120) $\\

{\multirow{3}{*}{155}} & {\multirow{3}{*}{R32}} & Q:$\; (\yb \xb \zb,\zz), \; (e,110)\; \textrm{X,B}: \; (\zb \yb \xb, \zz), \; (e, 1 0 1) \;  $ & $\textrm{P, P1, Q, Z1}: \; (y z x, \zz), \; (e, 120) $ \\ && $\textrm{P}: \; (x-y \,\xb \zb,\zz), \; (e, 100) \; \textrm{P1}:\;(\xb \, y-x \, \zb,\zz), (e,010) $ & \\ && $\textrm{B1}: (\xb \zb \yb, \zz), (e,011)$ & \\

{\multirow{2}{*}{160}}  & {\multirow{2}{*}{R3m}} & $\textrm{P}: \; (x z y, \zz), \; (e,011)\; \textrm{P1,P2}: \; (y x z, \zz), (e,110) \;$& - \\ && $\textrm{B,X}: \; (\yb \xb z,\zz), (e,110)$ & \\ 

161 & R3c & $\textrm{P}: \; (x z y, \oh \oh \oh) \; \textrm{P1,P2}: \; (y x z, \oh \oh \oh) $ &  - \\

{\multirow{3}{*}{166}} & {\multirow{3}{*}{R$\overline{\textrm{3}}$m}} & P:$(x z y,\zz), (e, 011)  \; \textrm{P1,P2}: (y x z, \zz), (e,110) \; $ & -\\ && X,B:$\;(\zb \yb \xb, \zz), (e,101) \;  \textrm{B1}: (\xb \zb \yb, \zz), (e,011)$ & \\
&& $\textrm{Q}: (\yb \xb \zb, \zz), (e,110)$ &\\

{\multirow{3}{*}{167}} & {\multirow{3}{*}{R$\overline{\textrm{3}}$c}} & P:$(x z y, \oh \oh \oh) \; \textrm{P1,P2}: (y x z, \oh \oh \oh)\; $ & -\\ && X,B:$\;(\zb \yb \xb, \oh \oh \oh), (e,101) \;  \textrm{B1}: (\xb \zb \yb, \oh \oh \oh), (e,011)$ & \\
&& $\textrm{Q}: (\yb \xb \zb, \zz), (e,110)$ &\\

\bottomrule
\end{tabular*}
\end{minipage}
\end{center}
\end{table}	 

\begin{table}[p]
\begin{center}
\begin{minipage}{\textwidth}
\caption{Generators for Abelian local symmetry groups of space groups with Cubic face-centred Bravais lattice. Plot of the corresponding Brillouin zone is reported in Fig.\ref{fig:cF}.} \label{tab:Gen_cF}
\begin{tabular*}{\textwidth}{@{\extracolsep{\fill}}rlllcc@{\extracolsep{\fill}}}
\toprule
\multicolumn{2}{@{}c@{}}{Space group} & Case 1 \\
\midrule

202	 &	Fm$\overline{\textrm{3}}$	& $\textrm{K}: \; \;(x y \zb, \zz) \; \textrm{U}: \; (x \yb z, \zz)$\\

203  & Fd3 & $\textrm{K}: \; (x y \zb, \tq \oq \tq), (e, 100) \; \textrm{U}: \; (x \yb z, \tq \tq \oq), (e, 001)$ \\

209  & F432& $\textrm{K}: \; (y x \zb,\zz), \; (e,110)\; \textrm{U}: \; (z \yb x, \zz), (e, 101)$ \\

210 & F4$_\textrm{1}$32 & $\textrm{K}: \; (y x \zb,\oq \oq \oq), (e,001)\; \textrm{U}: \; (z \yb x, \oq \tq \tq), (e, 010)$ \\

216  & F43m & $\textrm{K}: \; (y x z, \zz), \; (e, 110)\; \textrm{U}: (z y x, \zz), (e,101) $ \\

219 & F43c & $\textrm{K}: \; (y x z, \oh 0 0), (e,010) \; \textrm{U}: (z y x,0 0 \oh), (e,100)$ \\

225 & Fm3m & $\textrm{K}: \; (y x z,\zz), (y x \zb, \zz), (e,110) \; \textrm{U}: (x \yb z,\zz), (z \yb x, \zz), (e, 101) $\\

227 & Fd3m	& $\textrm{K}: \; (y x \zb, \oq \oq \oq) \; (e, 00\overline{1})$ \\

\bottomrule
\end{tabular*}
\end{minipage}
\end{center}
\end{table}	 

\end{document}